%% file: hff.tex
\documentclass[12pt]{article}

\usepackage{cite}
\usepackage{epsfig}

\input paperdef

\input draft

\begin{document}
\thispagestyle{empty}

\def\thefootnote{\fnsymbol{footnote}}

\begin{flushright}
CERN--TH/2000-056\\
DESY 00--028\\
KA-TP--9--1999\\
hep-ph/0003022 \\
\end{flushright}

\vspace{1cm}

\begin{center}

{\large\sc {\bf Decay widths of the neutral $\cp$-even MSSM Higgs bosons}}

\vspace{0.4cm}

{\large\sc {\bf in the Feynman-diagrammatic approach}}
 
\vspace{1cm}

{\sc 
S.~Heinemeyer$^{1}$%
\footnote{email: Sven.Heinemeyer@desy.de}%
, W.~Hollik$^{2}$%
\footnote{email: Wolfgang.Hollik@physik.uni-karlsruhe.de}%
 and G.~Weiglein$^{3}$%
\footnote{email: Georg.Weiglein@cern.ch}
}

\vspace*{1cm}

{\sl
$^1$ DESY Theorie, Notkestr. 85, 22603 Hamburg, Germany

\vspace*{0.4cm}

$^2$ Institut f\"ur Theoretische Physik, Universit\"at Karlsruhe, \\
D--76128 Karlsruhe, Germany

\vspace*{0.4cm}

$^3$ CERN, TH Division, CH-1211 Geneva 23, Switzerland
}

\end{center}

\vspace*{1cm}

\begin{abstract}
In the Minimal Supersymmetric Standard Model (MSSM) we incorporate the
Higgs-boson propagator corrections, evaluated up to \twol\ order,
into the prediction of $\Ga(\hff)$ and $BR(\hff)$ for $f = b, c, \tau$.
The propagator corrections consist of the full \onel\ contribution,
including the effects of non-vanishing external momentum, and
corrections of $\oaas$ 
at the \twol\ level.
The results are supplemented with the dominant \onel\ QED
corrections and final state QCD corrections from both gluons and gluinos.
The effects 
of the \twol\ propagator corrections and of the \onel\ gluino
contributions are 
investigated in detail. Our results are compared with the result obtained
within the renormalization group approach. Agreement within 10\% is found
for most parts of the MSSM parameter space.
\end{abstract}

\def\thefootnote{\arabic{footnote}}
\setcounter{page}{0}
\setcounter{footnote}{0}

\newpage


\section{Introduction}

The search for the lightest Higgs boson is a crucial test of 
Supersymmetry (SUSY) that can be performed with the present and the
next generation of accelerators. The prediction of a relatively
light Higgs boson is common to all supersymmetric models whose
couplings remain in the perturbative regime up to a very high energy
scale~\cite{susylighthiggs}. Finding the Higgs boson is thus one of
the main goals of todays high-energy physics.
Concerning the Higgs boson search, it is necessary to 
know the decay widths and branching ratios 
of the main decay channels to a high accuracy.
After the detection of a scalar particle it is mandatory as a next step
to measure its couplings to gauge bosons and fermions and also its
self-couplings very accurately, in order to establish the Higgs
mechanism and the Yukawa interactions experimentally. The determination
of the trilinear Higgs-boson self-couplings 
might be possible at a future linear \epem\ collider with high
luminosity~\cite{hhh}.

In this paper we concentrate on the coupling of the lightest MSSM
Higgs boson to Standard model (SM) fermions. In the MSSM the mass of
the lightest Higgs boson, $\mh$, is bounded from above by 
$\mh \lsim 135 \gev$, including radiative corrections up to \twol\
order~\cite{mhiggs1l,mhiggsf1l,mhiggsRG1a,mhiggsRG1b,mhiggsRG2,hoanghempfling,zhang,mhiggsletter,mhiggslong}.
Since the $b$-, the $c$-quark and the $\tau$-lepton are the heaviest
particles for which the decay $\hff$ is kinematically allowed, it is of
particular interest to calculate their corresponding decay rates and
branching ratios with high precision
\cite{hff1l,gammagluon1,gammagluon2,gammagluon3,gluon,gluongluino1,hffhigherorder}.
We analyze these decay rates and branching ratios, taking into account
the Higgs-boson propagator corrections, where at the
\onel\ level the full momentum dependence is kept. 
These corrections contain the Yukawa contributions of \order{\gf\mt^4/\MW^2},
which are the dominant electroweak \onel\ corrections to the Higgs-boson
decay width, and the corresponding QCD corrections of 
\order{\gf\als\mt^4/\MW^2}
as well as the Yukawa corrections of $\ogmzmtsomwz$. We also take into
account the \onel\ vertex corrections resulting from gluon, gluino
and photon exchange together with real gluon and photon emission as
given in \citere{hff1l}. 
Only the purely weak $\oa$ vertex corrections have been neglected.
We numerically investigate the effect of the \twol\ propagator
contributions and the
\onel\ gluino-exchange vertex correction. The latter one has been
mostly neglected in experimental analyses so far, but
can have a large impact on the result.
We show analytically that the Higgs-boson propagator correction with
neglected momentum dependence can be absorbed into the tree-level
coupling using the effective mixing angle from the neutral $\cp$-even
Higgs boson sector. The result in this approximation
is then compared with the full
result. We also perform a comparison between our full result and the
renormalization group (RG) improved effective potential calculation,
see \citere{hdecay} and references therein. 
For most parts of the MSSM parameter space we find agreement 
within 10\% between the two approaches, although deviations up to 50\%
are possible for certain ranges of the parameter space.

The paper is organized as follows: in section~2 we give the
calculational basis needed for the incorporation of the \twol\
Higgs-propagator corrections into the decay widths. We show
analytically how the propagator 
corrections with neglected external momentum are related to the
effective mixing-angle approach. 
The results for the gluon, gluino and QED vertex corrections in
combination with gluon and photon bremsstrahlung are
reviewed. In section~3 a numerical analysis for the decay rates and a
comparison of the 
full result and the effective mixing-angle result
is performed. Special emphasis is put on the \twol\ propagator
correction and the \onel\ gluino contribution.
Section~4 contains the comparison of our full result
with the RG approach. The conclusions can be found in section~5.


\section{Calculational basis}
\label{sec:calculbasis}

The  Higgs potential of the MSSM is given by~\cite{hhg}
\BEA
\label{Higgspot}
V &=& m_1^2 H_1\bar{H}_1 + m_2^2 H_2\bar{H}_2 - m_{12}^2 (\epsilon_{ab}
      H_1^aH_2^b + \hc)  \nonumber \\
   && \mbox{} + \frac{g'^2 + g^2}{8}\, (H_1\bar{H}_1 - H_2\bar{H}_2)^2
      +\frac{g^2}{2}\, |H_1\bar{H}_2|^2,
\EEA
where $m_1, m_2, m_{12}$ are soft SUSY-breaking terms, 
$g, g'$ are the $SU(2)$ and $U(1)$ gauge couplings, and 
$\epsilon_{12} = -1$.
The doublet fields $H_1$ and $H_2$ are decomposed  in the following way:
\BEA
H_1 &=& \VL H_1^1 \\ H_1^2 \VR = \VL v_1 + (\phi_1^{0} + i\chi_1^{0})
                                 /\sqrt2 \\ \phi_1^- \VR ,\non \\
H_2 &=& \VL H_2^1 \\ H_2^2 \VR =  \VL \phi_2^+ \\ v_2 + (\phi_2^0 
                                     + i\chi_2^0)/\sqrt2 \VR.
\label{eq:hidoubl}
\EEA
Besides $g$ and $g'$, two independent parameters enter the
potential~(\ref{Higgspot}):
$\Tb = v_2/v_1$ and $M_A^2 = -m_{12}^2(\Tb+\CTb)$,
where $M_A$ is the mass of the $\cp$-odd $A$ boson.

\medskip
The $\cp$-even neutral mass eigenstates are obtained performing the
rotation 
\BEA
\VL H^0 \\ h^0 \VR &=& \ML \Ca & \Sa \\ -\Sa & \Ca \MR 
\VL \phi_1^0 \\ \phi_2^0 \VR  
\equiv D^{-1}(\al) \VL \phi_1^0 \\ \phi_2^0 \VR
\label{higgsrotation}
\EEA
with the mixing angle $\alpha$ related to 
$\Tb$ and $M_A$ by
\BE
\tan 2\alpha = \tan 2\beta \frac{\MA^2 + \MZ^2}{\MA^2 - \MZ^2},
\quad - \frac{\pi}{2} < \alpha < 0 . 
\EE
At the tree level the mass matrix of the neutral $\cp$-even Higgs bosons
in the $\phi_1,\phi_2$ basis can be expressed  
in terms of $\MZ$ and $\MA$ as follows:
\BEA
M_{\rm Higgs}^{2, {\rm tree}} &=& \ML \mpe^2 & \mpez^2 \\ 
                           \mpez^2 & \mpz^2 \MR \non\\
&=& \ML \MA^2 \SQb + \MZ^2 \CQb & -(\MA^2 + \MZ^2) \Sb \Cb \\
    -(\MA^2 + \MZ^2) \Sb \Cb & \MA^2 \CQb + \MZ^2 \SQb \MR.
\EEA
Transforming to the eigenstate basis (\ref{higgsrotation}) yields 
\BE
M_{\rm Higgs}^{2, {\rm tree}} 
   \stackrel{\al}{\longrightarrow}
   \ML \mH^2 & 0 \\ 0 &  \mh^2 \MR
\EE
with $\mh$ and $\mH$ being the tree-level masses of the neutral
$\cp$-even Higgs bosons.

\bigskip
In the Feynman diagrammatic (FD) approach the higher-order corrected 
Higgs boson masses, denoted by $\Mh, \MH$, are derived by finding the
poles of the $h,H$-propagator 
matrix whose inverse 
is given by
\BE
\left(\Delta_{\rm Higgs}\right)^{-1}
= - i \ML q^2 -  \mH^2 + \hSi_{H}(q^2) &  \hSi_{hH}(q^2) \\
     \hSi_{hH}(q^2) & q^2 -  \mh^2 + \hSi_{h}(q^2) \MR,
\label{higgsmassmatrixnondiag}
\EE
where the $\hSi(q^2)$ denote the 
renormalized Higgs boson self-energies. For these self-energies
we take the result up to \twol\ order, see \refse{subsec:evaloaas} below.

\bigskip
Our main emphasis in this paper is on the fermionic decays of the
light Higgs boson, but for completeness we list the expressions for both
$h$ and $H$.
The amplitudes for the decays $h,H \rightarrow f\bar{f}$
can be written as follows:

\BEA
\label{ewdecayamplitude}
A(\hff) &=& \wZh \KL \Gh + \ZhH\; \GH \KR~, \\
A(\Hff) &=& \wZH \KL \GH + \ZHh\; \Gh \KR~, 
\EEA
with
\BEA
\label{ZhH}
\ZhH &=& - \frac{\hSihH(\Mh^2)}
                {\Mh^2 - \mH^2 + \hSiH(\Mh^2)}, \\
\label{ZHh}
\ZHh &=& - \frac{\hSihH(\MH^2)}
                {\MH^2 - \mh^2 + \hSih(\MH^2)},
\EEA
involving the renormalized self-energies $\hSi(q^2)$ and 3-point vertex
functions $\Gh, \GH$. 
The wave function renormalization factors $\Zh$ and
$\ZH$  are related to the finite residues of the $h$ and $H$ propagators,
respectively:
\BEA
\label{zlh}
\Zh &=& 
        \ed{1 + \re\hSiph(q^2) - 
        \re\KL \frac{\hSihH^2(q^2)}
            {q^2 - \mH^2 + \hSiH(q^2)} \KR '}~_{\Bigr| q^2 = \Mh^2} \\
\label{zhh}
\ZH &=& 
        \ed{1 + \re\hSipH(q^2) - 
        \re\KL \frac{\hSihH^2(q^2)}
            {q^2 - \mh^2 + \hSih(q^2)} \KR '}~_{\Bigr| q^2 = \MH^2}~.
\EEA


\subsection{The $\aeff$-approximation}
\label{subsec:aeff}


The dominant contributions for the Higgs boson self-energies can be
obtained by setting $q^2=0$. Approximating the 
renormalized Higgs boson self-energies by
\BE
\hSi(q^2) \to \hSi(0) \equiv \hSi
\label{zeroexternalmomentum}
\EE
yields the
Higgs boson masses by re-diagonalizing the dressed
mass matrix
\BE
\label{deltaalpha}
M_{\rm Higgs}^{2} 
 = \ML \mH^2 - \hSiH & -\hSihH \\
       -\hSihH & \mh^2 - \hSih \MR
   \stackrel{\De\al}{\longrightarrow}
   \ML \MH^2 & 0 \\ 0 &  \Mh^2 \MR ,
\EE
where $\Mh$ and $\MH$ are the corresponding
higher-order-corrected Higgs boson
masses. The rotation matrix in the transformation~(\ref{deltaalpha}) 
reads:
\BE
D(\De\al) = \ML \CDea & -\SDea \\ \SDea & \CDea \MR~.
\label{rotmatrix}
\EE
The angle $\De\al$ is related to the renormalized
self-energies and masses through the eigenvector equation
\BEA
&& 
\ML \mH^2 - \hSiH - \Mh^2 & -\hSihH \\
    -\hSihH & \mh^2 - \hSih - \Mh^2 \MR
\VL -\SDea \\ \CDea \VR~=~0 
\EEA
which yields
\BE
\frac{\hSihH}{\Mh^2 - \mH^2 + \hSiH}~=~\TDea~.
\label{tandeltaalpha}
\EE
The second eigenvector equation leads to:
\BE
\frac{-\hSihH}{\MH^2 - \mh^2 + \hSih}~=~\TDea~.
\EE

\noindent
With the approximation~(\ref{zeroexternalmomentum}) one deduces
\BEA
\label{ZhHTanDeltaalpha}
\ZhH &=& - \frac{\hSihH}{\Mh^2 - \mH^2 + \hSiH} = - \TDea, \\
\label{ZHhTanDeltaalpha}
\ZHh &=& - \frac{\hSihH}{\MH^2 - \mh^2 + \hSih} = + \TDea,
\EEA
and $\Zh$ can be expressed as 
\BEA
\Zh &=&
        \ed{1 + \KL \frac{\hSihH}{\Mh^2 - \mH^2 + \hSiH} \KR^2} \non \\
    &=& \ed{1 + \TQDea} = \CQDea~.
\EEA
Analogously one obtains 
\BE
\ZH = \CQDea~.  
\EE

\bigskip \noindent
At the tree level, the vertex functions can be written as
\BEA
\renewcommand{\arraystretch}{1.5}
\left. \begin{array}{l}
\Gh = \frac{i e \mf \Sa}{2\sw\MW\Cb} = \cfpfp \; \Sa  \\
\GH = \frac{-i e \mf \Ca}{2\sw\MW\Cb} = - \cfpfp \; \Ca 
\end{array} \KKKR \mbox{for d-type fermions} \\
\renewcommand{\arraystretch}{1.5}
\left. \begin{array}{l}
\Gh = \frac{-i e \mf \Ca}{2\sw\MW\Sb} = \cff \; \Ca,  \\
\GH = \frac{-i e \mf \Sa}{2\sw\MW\Sb} = \cff \; \Sa 
\end{array} \KKKR \mbox{for u-type fermions}~.
\renewcommand{\arraystretch}{1}
\EEA
Incorporating them into the decay amplitude yields:
\BEA
A_{\rm eff}(\hff) &=& \wZh \KL \Gh + \ZhH\; \GH \KR \non \\
 &=& \cfpfp \CDea \KL \Sa - \TDea \KL - \Ca \KR \KR \non \\
 &=& \cfpfp \sin(\al + \De\al) \non \\
\label{ampeffhbb}
 &\equiv& \cfpfp \Saeff~~~\mbox{(for d-type fermions)} \\
 A_{\rm eff}(\hff) &\equiv& \cff \Caeff~~~\mbox{(for u-type fermions)} \\
 A_{\rm eff}(\Hff) &\equiv& -\cfpfp \Caeff~~~\mbox{(for d-type fermions)} \\
 A_{\rm eff}(\Hff) &\equiv& \cff \Saeff~~~\mbox{(for u-type fermions)}
\EEA

\smallskip
\noindent
Recalling the relations
\BE
D(\aeff) = D(\al)\;D(\De\al)
\EE
and
\BE
\ML \hSiH & \hSihH \\ \hSihH & \hSih \MR = 
D^{-1}(\al) \ML \hSi_{\Pe} & \hSi_{\PePz} \\ 
                \hSi_{\PePz} & \hSi_{\Pz} \MR D(\al)
\EE
it is obvious that $\aeff = (\al + \De\al$) is exactly the angle that
diagonalizes the higher order corrected Higgs boson mass matrix in the
$\Pe,\Pz$-basis:
\BEA
\label{higgsmassmatrixPhi1Phi2}
&&
\ML \mpe^2 - \hSi_{\Pe} & \mpez^2 - \hSi_{\PePz} \\ 
    \mpez^2 - \hSi_{\PePz} & \mpz^2 - \hSi_{\Pz} \MR 
   \stackrel{\aeff}{\longrightarrow}
   \ML \MH^2 & 0 \\ 0 &  \Mh^2 \MR~ \non \\
&& \Big\downarrow~~\al \\
&&
\ML \mH^2 - \hSiH & - \hSihH \\ 
    - \hSihH & \mh^2 - \hSih \MR 
 \stackrel{\De\al}{\longrightarrow}
   \ML \MH^2 & 0 \\ 0 &  \Mh^2 \MR . \non
\EEA
$\aeff$ can be obtained from 
\BE
\aeff = {\rm arctan}\KKL 
  \frac{-(\MA^2 + \MZ^2) \Sb \Cb - \hSi_{\PePz}}
       {\MZ^2 \CQb + \MA^2 \SQb - \hSi_{\Pe} - \mh^2} \KKR~, ~~
 -\frac{\pi}{2} < \aeff < \frac{\pi}{2}~.
\EE


\subsection{The Higgs-boson propagator corrections}
\label{subsec:evaloaas}

For the Higgs boson self-energies employed in
\refeqs{higgsmassmatrixnondiag}--(\ref{zhh}) we use the currently
most accurate result
based on Feynman-diagrammatic calculations. It contains the result of
the complete \onel\ on-shell calculation of \citere{mhiggsf1l},
together with the dominant \twol\ corrections of
$\oaas$ obtained in \citeres{mhiggsletter,mhiggslong}, including also the
leading terms of $\ogmzmtsomwz$~\cite{mhiggsRG1a,mhiggsRG1b,mhiggsRG2};
the Fortran program \fh, based on  
this result, has been described in \citere{feynhiggs}. In this way the
complete  MSSM \onel\ on-shell result together with the
dominant \twol\ contribution, originating from the $t-\Stop$-sector
(without any restrictions 
on the mixing), is taken into account in the propagator corrections.

In the approach in \citeres{mhiggsletter,mhiggslong} 
the Higgs boson self-energies are given by:
\BE
\hSi_s(q^2) = \hSie_s(q^2) + \hSiz_s(0)~,~~s = h, H, hH~,
\label{hsefull}
\EE
where the momentum dependence has been neglected only at the \twol\ level,
while the full momentum dependence is kept in the \onel\
contributions.

In a first step of approximation
for the calculation of the decay width $\Ga(\hff)$ 
the momentum dependence is
neglected everywhere in the Higgs boson self-energies (see
\refeq{zeroexternalmomentum}):
\BE
\hSi_s(q^2) \to \hSie_s(0) + \hSiz_s(0)~,~~s = h, H, hH~.
\label{hseq2zero}
\EE
This corresponds to the $\aeff$-approximation, as described in
\refse{subsec:aeff}.

In a second step of approximation 
we approximate the Higgs boson self-energies by
the compact analytical formulas given in \citere{mhiggslle}:
\BE
\hSi_s(q^2) = \hSi_s^{(1)\rm\, approx}(0) 
            + \hSi_s^{(2)\rm\, approx}(0)~,~~s = h, H, hH ,
\label{hselle}
\EE
yielding relatively short expressions which allow a very fast
numerical evaluation. In the following, this approximation
is labeled by $\aeffapprox$.

\bigskip
For the $\Stop$-sector, we use the same
conventions as in~\citere{mhiggslong}:
the scalar top masses and the mixing angle are related to the
parameters $M_{{\tilde t}_L}$, $M_{{\tilde t}_R}$ and $\Xt$ of the 
$\Stop$-mass matrix
\BE
\label{stopmassenmatrix}
{\cal M}^2_{\Stop} = 
  \ML \MstL^2 + \mt^2 + \CZb (\edz - \frac{2}{3} \sw^2) \MZ^2 &
      \mt \Xt \\
      \mt \Xt &
      \MstR^2 + \mt^2 + \frac{2}{3} \CZb \sw^2 \MZ^2 
  \MR~,
\EE
with
\BE
\Xt = A_t - \mu \CTb~.
\label{mtlr}
\EE
In the numerical analysis below we have chosen 
$\msq \equiv \MstL = \MstR$.

In \citeres{mhiggslong,mhiggslle} it has been shown that for a given
set of MSSM parameters the maximal values of $\Mh$ as a function of
$\Xt$ are obtained for 
$|\Xt/\msq| \approx 2$. This case we refer to as
'maximal mixing'. Minimal values for $\Mh$ are reached for 
$\Xt \approx 0$. This 
case we refer to as 'no mixing'.


\subsection{Decay width of the lightest Higgs boson}
\label{subsec:hdecaywidth}

At the tree level, the decay width for $h\rightarrow f\bar{f}$
is given by
\BE
\label{decaywidthtree}
\Gz(\hff) = N_C \frac{\mh}{8\,\pi}
           \KL 1 - \frac{4\,\mf^2}{\mh^2} \KR^\frac{3}{2}~|\Gh|^2~.
\EE
The electroweak propagator corrections 
are incorporated by using the higher-order decay
amplitude~(\ref{ewdecayamplitude})
\BE
\label{decaywidthoneloopew}
\Ge \equiv \Ge(\hff) =  N_C \frac{\Mh}{8\,\pi}
            \KL 1 - \frac{4\,\mf^2}{\Mh^2} \KR^\frac{3}{2}~|A(\hff)|^2~.
\EE
The $\aeff$-approximation is given by
\BE
\Geeff \equiv \Geeff(\hff) = N_C \frac{\Mh}{8\,\pi}
       \KL 1 - \frac{4\,\mf^2}{\Mh^2} \KR^\frac{3}{2}~|A_{\rm eff}(\hff)|^2~.
\label{gameff}
\EE
In this paper we consider only those electroweak higher-order contributions
which enter via the Higgs boson self-energies. 
These corrections contain the Yukawa contributions of \order{\gf\mt^4/\MW^2},
which are the dominant electroweak \onel\ corrections to the Higgs-boson
decay width, and the corresponding dominant \twol\ corrections, see
\refse{subsec:evaloaas}. 
The pure weak $\oa$
vertex corrections are neglected (they have been calculated in
\citere{hff1l} and were found to be 
at the level of only a few \%
for most parts of the MSSM parameter space, see also
\refse{subsubsec:decaywidthgluino} below.) 


\subsubsection{QED corrections}
\label{subsubsec:decaywidthgamma}

Here we follow the results given in
\citeres{gammagluon1,gammagluon2,gammagluon3,hff1l}.
The IR-divergent virtual photon contribution is taken into account
in combination with real-photon bremsstrahlung yielding the QED 
corrections. The contribution to the decay width induced by
$\ga$-exchange and final-state photon radiation can be cast into the
very compact formula
\BE
\De\Gga = \Ge\cdot\deqed~,
\EE
where for $\mf^2 \ll \Mh^2$ the factor $\deqed$ has the simple form
\BE
\deqed = \frac{\al}{\pi}Q_f^2 \KKL - 3 \log \KL \frac{\Mh}{\mf} \KR
                                   + \frac{9}{4} \KKR~.
\EE


\subsubsection{QCD corrections: gluon contributions}
\label{subsubsec:decaywidthgluon}

The corresponding results have been obtained in 
\citeres{gammagluon1,gammagluon2,gammagluon3,gluon,gluongluino1,hff1l,hffhigherorder}.
The additional contribution to the decay width induced by
gluon exchange and final-state gluon radiation can be incorporated
into~(\ref{decaywidthoneloopew}) by writing
\BE
\label{decaywidthgluon}
\Geg = \Ge \cdot \frac{\mq^2(\Mh^2)}{\mq^2}
       \KKL 1 + \frac{\als(\Mh^2)}{\pi} 
  \KKKL \cf \frac{9}{4} 
    + \frac{8}{3} \KL 1 - \frac{\als(\mb^2)}{\als(\Mh^2)} \KR \KKKR \KKR~.
\EE
The correction factor containing $\als(\mb^2)$, which has not been included in
previous diagrammatic calculations, can give rise to non-negligible
contributions. $\mq(\Mh^2)$ 
is calculated via
\BEA
\mq(q^2) &=& \mq \frac{c(q^2)}{c(\mq^2)}~,\\
c(q^2) &=& \KL \frac{\ben\als(q^2)}{2\,\pi}\KR^{-\gan/2\ben}
\Bigg[ 1 + \frac{\KL \bee\gan - \ben\gae \KR}{\ben^2}
           \frac{\als(q^2)}{8\,\pi} \non \\
 && + \KL \frac{(\bee \gan - \ben \gae)^2}{2 \ben^4}
          + \frac{\gan (\bez \ben - \bee^2)}{\ben^3}
          + \frac{\gae \bee}{\ben^2} - \frac{\gaz}{\ben} \KR
          \KL \frac{\als(q^2)}{8\,\pi} \KR^2 \Bigg]~,
\label{cfkt}
\EEA
where $\mq$ is the pole mass and $\mq(\mq^2) = \mq$. The coefficients in
\refeq{cfkt} are:
\BEA
\ben &=& \frac{33 - 2\nf}{3}~, \non \\
\bee &=& 102 - \frac{38}{3}\nf~, \non \\
\bez &=& \frac{2857}{2} - \frac{5033}{18}\nf +
                          \frac{325}{54}\nf^2~, \non \\
\gan &=& -8~, \non \\
\gae &=& -\frac{404}{3} + \frac{40}{9}\nf~, \non \\
\gaz &=& \frac{2}{3} \KKL \frac{140}{27} \nf^2 
                         + \KL 160 \zeta(3) + \frac{2216}{9}\KR \nf
                         - 3747 \KKR~,
\EEA
where $\zeta(3) \approx 1.2020596\dots$ and $\nf = 5$ for 
$f = c, b$, which are considered here.
The strong coupling constant $\als$ is given up to three loops by:
\BE
\als(q^2) = \frac{4\,\pi}{\ben L_q}
  \KKL 1 - \frac{\bee}{\ben^2}\frac{\log L_q}{L_q} 
         + \frac{\bee^2}{\ben^4}\frac{\log^2 L_q}{L_q^2}
         - \frac{\bee^2}{\ben^4}\frac{\log L_q}{L_q^2}
         + \frac{\bez \ben - \bee^2}{\ben^4} \ed{L_q^2} \KKR~,
\EE
where $L_q = \log(q^2/\Laqcd^2)$. (For the numerical evaluation 
$\Laqcd = 220 \mev$ has been used.)
Numerically, more than 80\% of the gluon-exchange contribution is
absorbed into the running quark mass.


\subsubsection{QCD corrections: gluino contributions}
\label{subsubsec:decaywidthgluino}

We follow the calculation given in \citere{hff1l}%
\footnote{
An error in \citere{hff1l} concerning the proper inclusion of the
$Hf\bar{f}$ coupling has been corrected.
}%
, similar
results can also be found in \citere{gluongluino1}.
The additional contributions to the decay width induced by
gluino-exchange are incorporated via
\BE
\De\Ggl = \Ge \cdot \de\Ggl~,
\EE
where $\de\Ggl$ is given by
\BE
\label{decaywidthgluino}
\de\Ggl = \frac{2}{\Gh + \ZhH \GH}
 \re \KKL \Ggl^{h} + \ZhH \Ggl^{H} \KKR
\EE
for real $\ZhH$ (i.e.\ neglecting the imaginary part in
\refeq{decaywidthgluino});
$\Ggl^h$ and $\Ggl^H$ are given by
\BEA
\Ggl^h &=& \Gh \KKL \De T_{\gl}^h~_{\Bigr| q^2 = \Mh^2}
         + \Si^f_{S,\gl}(\mq^2) 
         - 2 \mq^2 \KL \Si_{S,\gl}^{f\prime}(\mq^2)
                     + \Si_{V,\gl}^{f\prime}(\mq^2) \KR \KKR \\
 && \non \\
\Ggl^H &=& \GH \KKL \De T_{\gl}^H~_{\Bigr| q^2 = \Mh^2}
         + \Si^f_{S,\gl}(\mq^2) 
         - 2 \mq^2 \KL \Si_{S,\gl}^{f\prime}(\mq^2)
                     + \Si_{V,\gl}^{f\prime}(\mq^2) \KR\KKR~. \\
 && \non
\EEA
$\De T_{\gl}^{h,H}$ denote the gluino vertex-corrections, whereas 
$\Si^f$ represents the gluino contribution to the fermion self-energy
corrections. Explicit expressions for these terms can be found
in~\citere{hff1l}.

\bigskip
For large values of $\tb$ in combination with large values of  $|\mu|$,
the gluino-exchange corrections to 
$\Ga(\hbb)$ can become very large.
In \citeres{gluinoresum1,higgssearch2} as well as in
\citere{gluinoresum2} it has been proposed to derive an effective
contribution to the decay width resummed to all orders.
A similar
resummation can be applied for weak $\oa$ chargino-exchange
corrections to $\Ga(\hbb)$, where they become
non-negligible~\cite{gluinoresum1}. 
In the numerical examples given in \citere{gluinoresum2} the
difference between the \onel\ result for $\Ga(\hff)$ and the effectively
resummed result does not exceed 10 -- 15\%, even for very large values
of $\tb > 40$. 
A proof of how the resummation of the leading terms arising for large
$\tb$ can be performed for the $H^+\bar{t}b$ vertex is given in
\citere{Hptbresum}.
Concerning our numerical analysis in \refse{sec:numanal} we have
neglected the additional contributions from the resummation. 
These additional corrections, although potentally large in some regions
of the MSSM parameter space, would not qualitatively change our
conclusions given below. A more detailed investigation of the effects of
a proper resummation of the leading contributions and of the inclusion
of the complete electroweak one-loop vertex corrections will be given in
a forthcoming publication.

For some parameter combinations the gluino corrections can drive
$\Ga(\hbb)$ to very small values, see the discussion 
at the end of \refse{subsec:twoloopeffect}.


\subsubsection{Decay width and branching ratio}

Including the various types of corrections, the decay width
is given by
\BE
\label{decaywidthfull}
\Ga(\hff) = \Geg + \De\Gga + \De\Ggl~.
\EE
Summing over $f = b, c, \tau$ and adding $\Ga(h \to gg)$ 
(which can be numerically relevant~\cite{hgg}),
results in an approximation for the total decay width
\BE
\Gtot = \sum_{f=b,c,\tau} \Ga(\hff) + \Ga(h \to gg)~.
\EE
We do not take into account the decay $h \to AA$ 
(see e.g.~\citere{haa} for a detailed study).
Although it is
dominant whenever it is kinematically allowed, 
it plays a role only for very small values of $\tb$ ($\tb \lsim 1.5$)
which will not be considered here because of the limits obtained at
LEP2~\cite{tbexcl}. We also assume that all other 
SUSY particles are too heavy to allow further decay channels.
In addition, we neglect the decay $h \to WW^*$ which can be of 
${\cal O}(1\%)$ for $\Mh \gsim 100 \gev$.

The fermionic branching ratio is defined by
\BE
\rf \equiv BR(\hff) = \frac{\Ga(\hff)}{\Gtot}~.
\EE


\section{Numerical analysis}
\label{sec:numanal}

Concerning the numerical evaluation of the Higgs-boson propagator
corrections, we follow \refse{subsec:evaloaas}. For
$\tb$ we have chosen two representative values,
a relatively low value, $\Tb = 3$,%
\footnote{
This values is well above the expected limit obtainable at the end of
LEP, assuming that no Higgs boson signal will be
found~\cite{tbexcl}. For these expected limits $\mt = 174.3 \gev$ and
$\msq = 1 \tev$ has been assumed.
}%
~and a high value, $\Tb = 40$.
For sake of comparison we also consider 
an intermediate value of $\Tb = 20$ in some cases.
If not indicated differently, the other MSSM parameters
are chosen as follows:
$\mu = -100 \gev$, $M_2 = \msq$ ($M_2$ is the
soft SUSY-breaking term in the gaugino sector), gluino mass 
$\mgl = 500 \gev$, $\Ab = \At$ (which fixes, together with $\mu$, 
the mixing in the $\Sbot$-sector). For the SM fermion masses
we have furthermore chosen $\mt = 175 \gev$, $\mb = 4.5 \gev$%
\footnote{
The value of $\mh = 4.5 \gev$ is used in the tree-level expression and
in the QED and QCD vertex corrections (see \refses{subsubsec:decaywidthgamma} -
\ref{subsubsec:decaywidthgluino}), while for the Higgs-propagator
corrections the running bottom mass, $\mb(\mt) = 2.97 \gev$, has been
used, in order to partially absorb higher-order QCD corrections.
}
 , $m_\tau = 1.777 \gev$ and $m_c = 1.5 \gev$.

The mass $\MA$
of the $\cp$-odd Higgs boson is treated as an input
parameter and is varied in the interval $50 \gev \le \MA \le 500 \gev$. 
The corresponding values for $\Mh$ follow from 
\refeq{higgsmassmatrixnondiag}. 
$\Mh$, derived in this way, subsequently enters
the numerical evaluation of the formulas presented in
section~\ref{sec:calculbasis}. Thus the variation of
$\Mh$ in the plots stems from the variation of $\MA$ in the above
given range.


\subsection{Effects of the \twol\ Higgs-propagator corrections}
\label{subsec:twoloopeffect}

\begin{figure}[ht!]
\vspace{1em}
\begin{center}
\mbox{
\psfig{figure=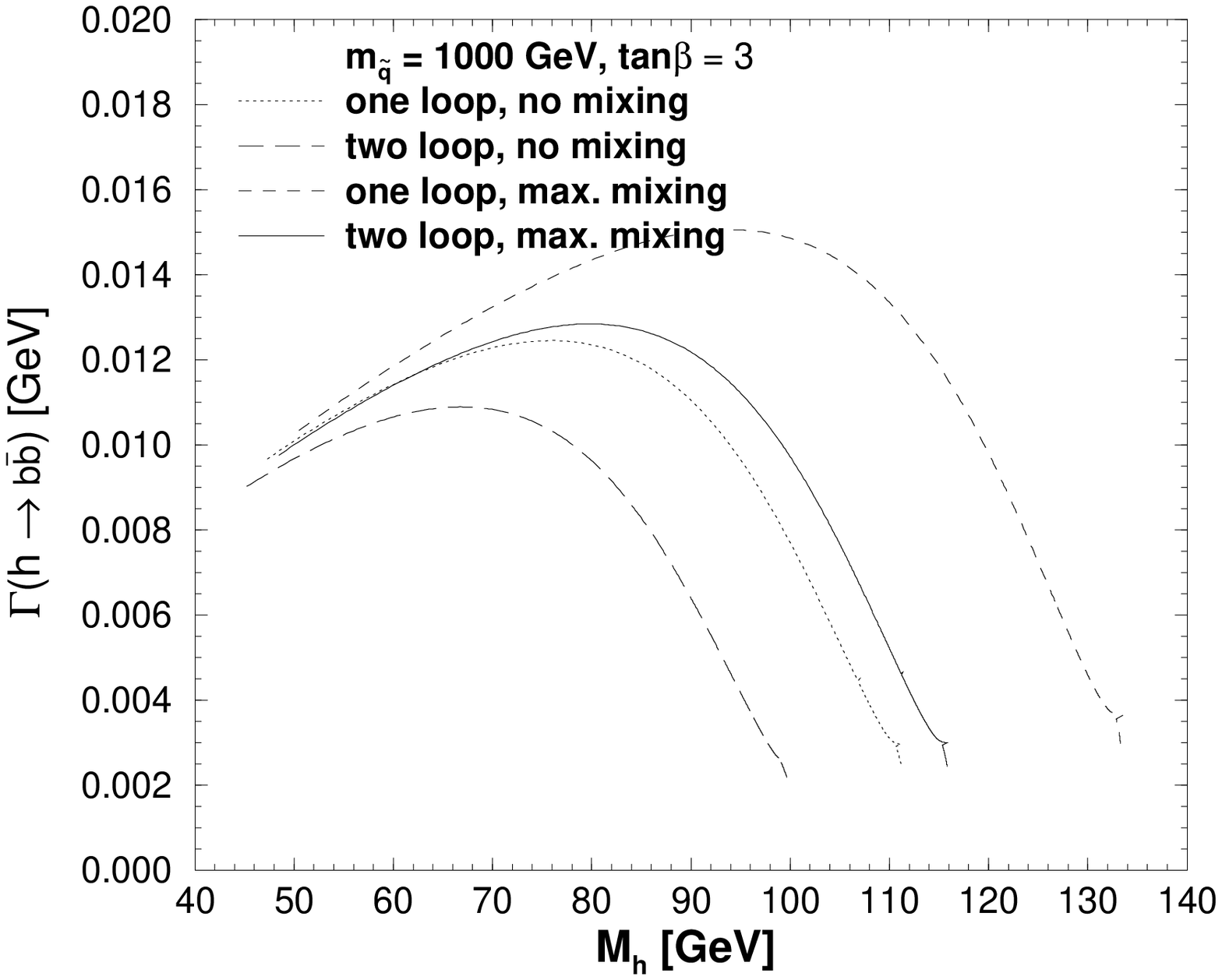,width=7cm,height=7cm}}
\hspace{1.5em}
\mbox{
\psfig{figure=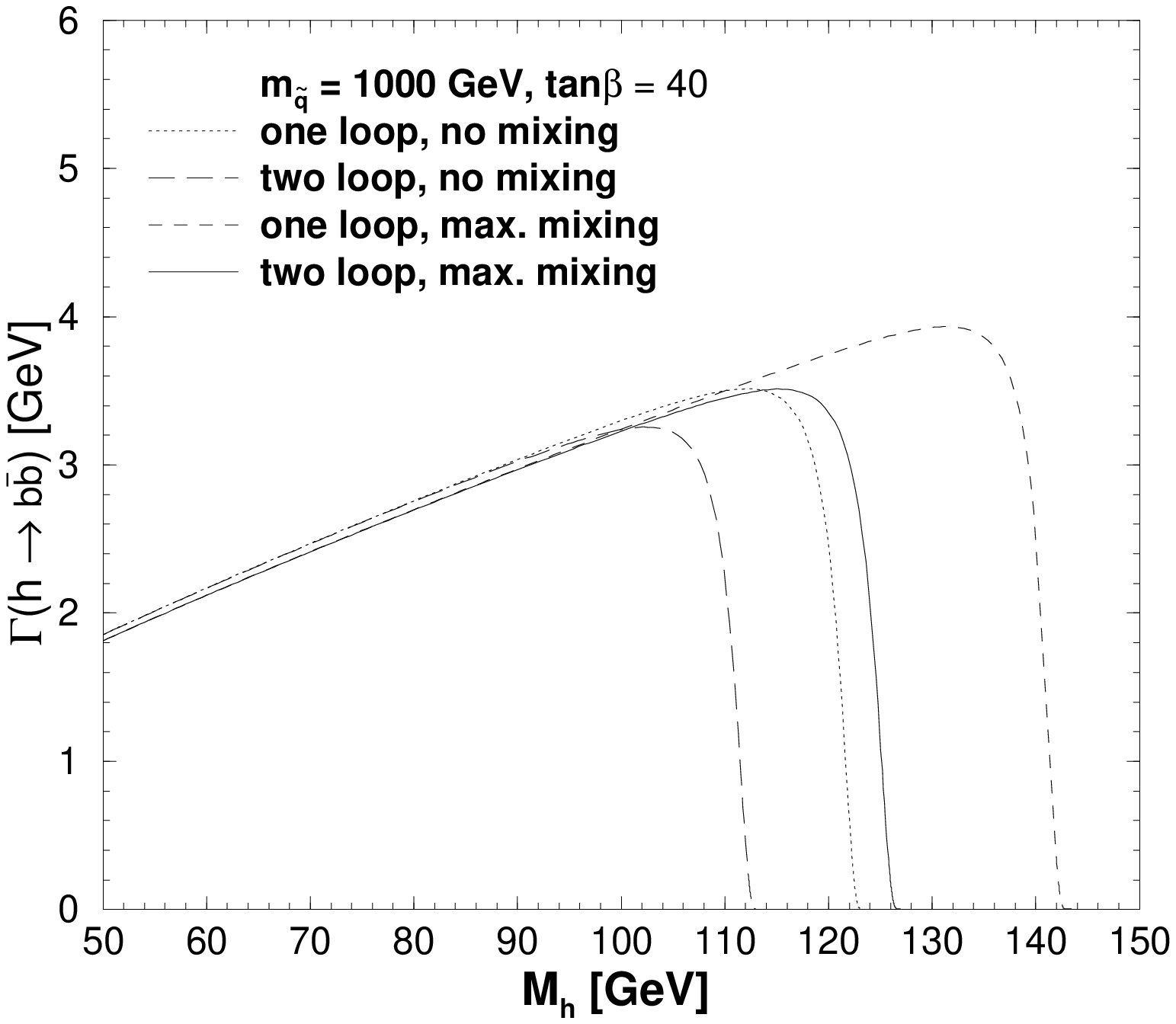,width=7cm,height=7cm}}
\end{center}
\caption[]{{\it
$\Ga(\hbb)$ is shown as a function of $\Mh$. The Higgs-propagator
corrections have 
been evaluated at the one- and at the \twol\ level. The QED, 
gluon  and gluino contributions are included. The other parameters are
$\mu = -100 \gev$, $M_2 = \msq$, $\mgl = 500 \gev$, $\Ab = \At$, 
$\Tb = 3, 40$. The result is given in the no-mixing and maximal-mixing
scenario.
}}
\label{fig:Ghbb}
\end{figure}

We first focus on the effects of the \twol\ Higgs-boson propagator
corrections. They have
been evaluated at the one- and at the \twol\ level as described in
\refse{subsec:evaloaas}. 
\reffi{fig:Ghbb} shows the 
results for $\Ga(\hbb)$ for a common scalar quark mass 
$\msq = 1000 \gev$ and $\Tb = 3$ and $\Tb = 40$ in the no-mixing and
the maximal-mixing scenario. 
The QED and the QCD
gluon and gluino vertex contributions are also included.

In the small $\tb$ scenario, larger values for $\Ga(\hbb)$ are obtained
for maximal mixing. The \twol\ corrections strongly reduce the decay width.
In the large $\tb$ scenario 
the variation is mainly a kinematical effect from the different values
of $\Mh$ at the one- and \twol\ level.
The absolute values obtained for $\Ga(\hbb)$ are three
orders of magnitude higher in the $\Tb = 40$ scenario, which is due
to the fact that $\Ga(\hbb) \sim 1/\CQb$. 

\bigskip
In \reffi{fig:Ghbbttcc} the three decay rates $\Ga(\hbb)$,
$\Ga(\htautau)$ and $\Ga(\hcc)$ are shown as a function of $\Mh$.
The results are given in the no-mixing scenario for $\msq = 500 \gev$
and $\Tb = 3, 40$. 

\begin{figure}[ht!]
\vspace{1em}
\begin{center}
\mbox{
\psfig{figure=gh2l08.bw.eps,width=7cm,height=7cm}}
\hspace{1.5em}
\mbox{
\psfig{figure=gh2l09.bw.eps,width=7cm,height=7cm}}
\end{center}
\caption[]{{\it
$\Ga(\hbb)$, $\Ga(\htautau)$ and $\Ga(\hcc)$ are shown as a function
of $\Mh$. The Higgs-propagator corrections have 
been evaluated at the one- and at the \twol\ level. The QED, 
gluon  and gluino contributions are included. The other parameters are
$\mu = -100 \gev$, $M_2 = \msq$, $\mgl = 500 \gev$, $\Ab = \At$, 
$\Tb = 3, 40$. The result is given in the no-mixing
scenario.
}}
\label{fig:Ghbbttcc}
\end{figure}

\begin{figure}[ht!]
\begin{center}
\mbox{
\psfig{figure=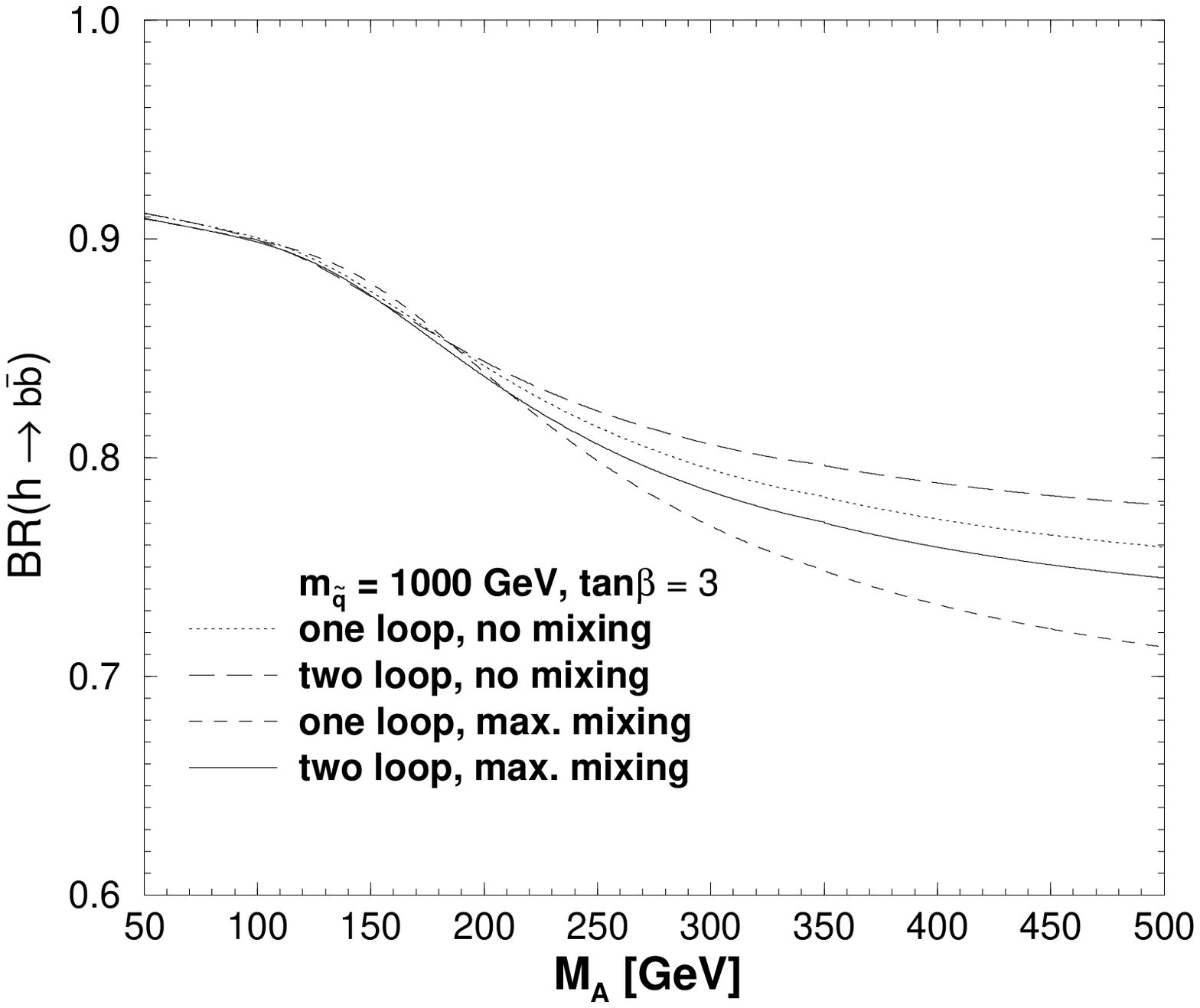,width=7cm,height=7cm}}
\hspace{1.5em}
\mbox{
\psfig{figure=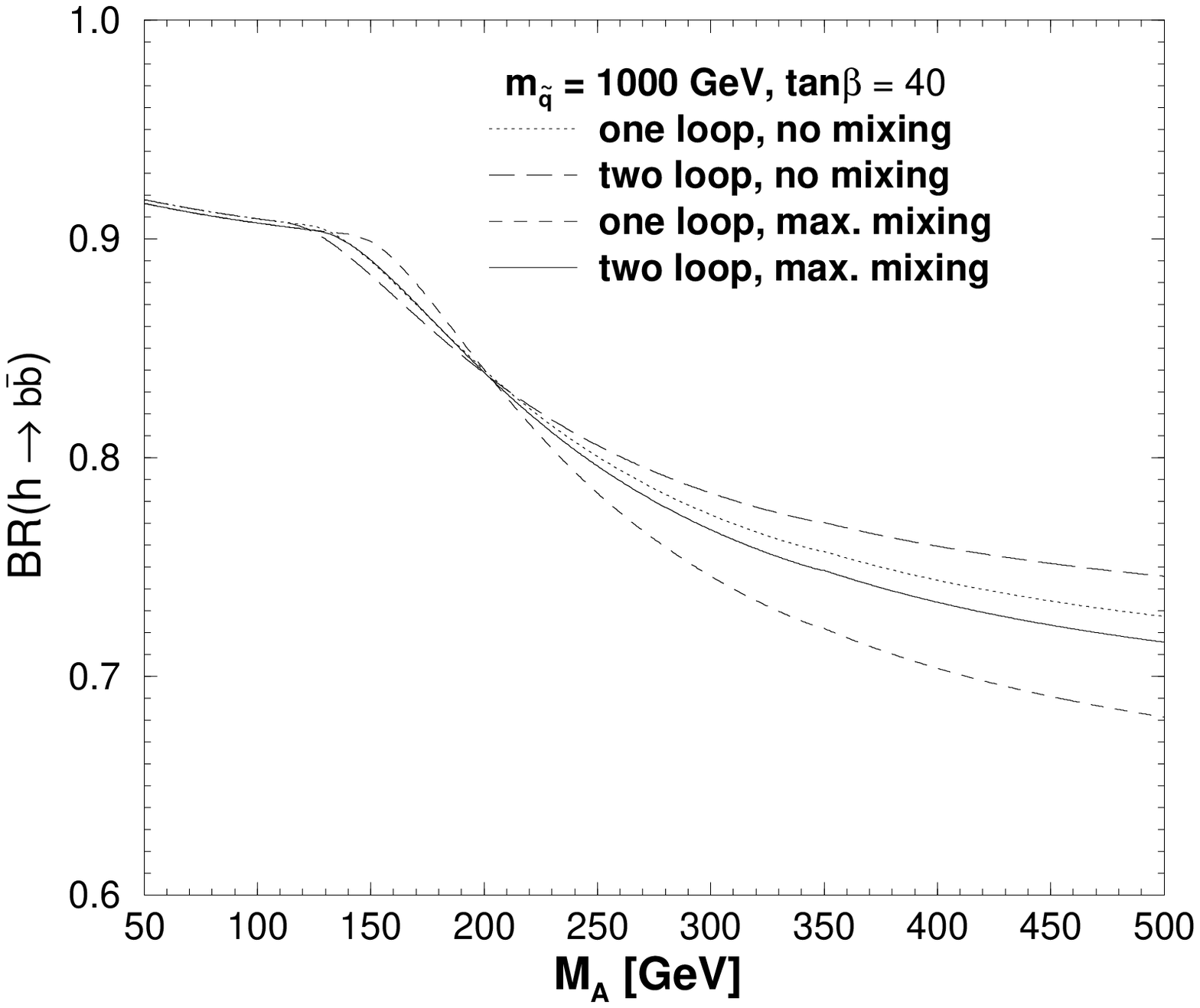,width=7cm,height=7cm}}
\end{center}
\begin{center}
\mbox{
\psfig{figure=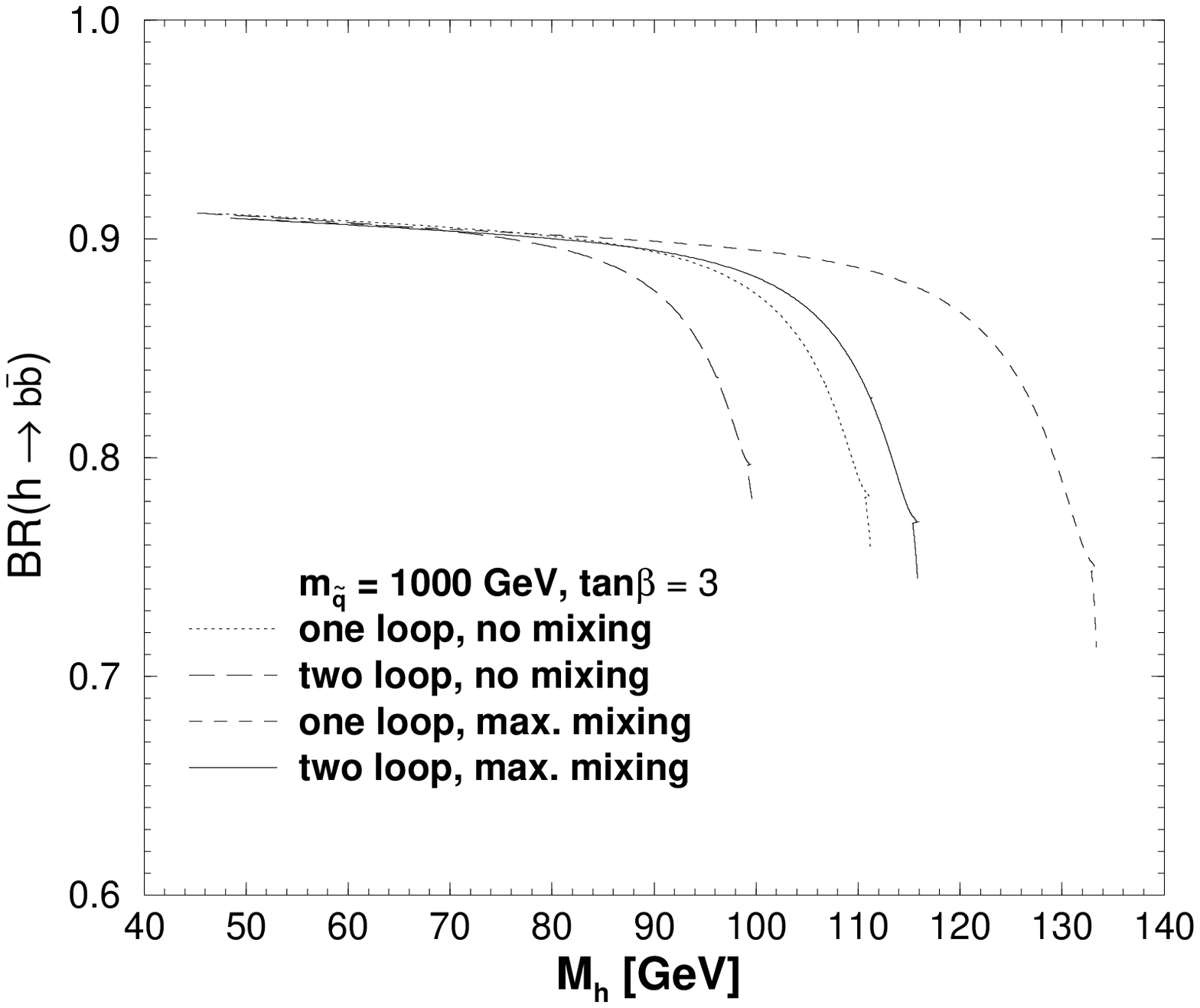,width=7cm,height=7cm}}
\hspace{1.5em}
\mbox{
\psfig{figure=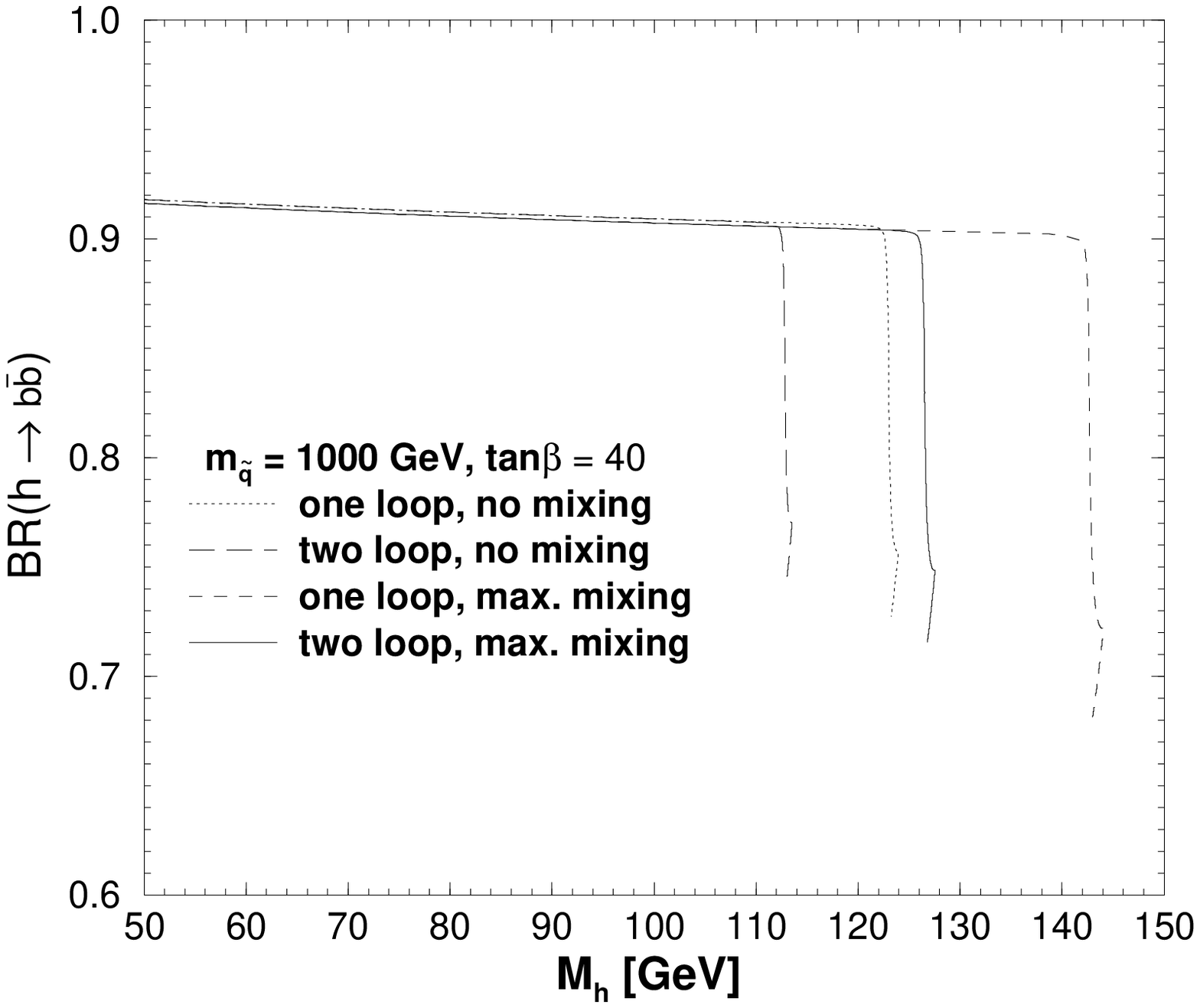,width=7cm,height=7cm}}
\end{center}
\caption[]{{\it
$BR(\hbb)$ is shown as a function of $\MA$ and $\Mh$ 
for the same settings as in
\reffi{fig:Ghbb}. The QED, gluon and gluino contributions are included.
}}
\label{fig:BRhbb}
\end{figure}

In the low $\tb$ scenario
$\Ga(\hbb)$ and $\Ga(\htautau)$ are lowered at the \twol\ level, while
$\Ga(\hcc)$ is increased. The decay rate for $\hbb$ is about one and
two orders of magnitude larger compared to the ones of $\htautau$ and
$\hcc$, respectively. In the large $\tb$ scenario the shifts are
again dominated by the kinematical effect from the different values
of $\Mh$ at the one- and \twol\ level.
In the maximal-mixing case, which is not plotted here, we find
qualitatively the same behavior. 

\smallskip
We now turn to the effects of the \twol\ corrections to the branching
ratios. 
For not too large values of $\Mh$, $\Gtot$ is strongly dominated
by $\Ga(\hbb)$. For large values of $\Mh$ the decay into gluons
becomes more relevant. 
In \reffi{fig:BRhbb} we show the branching ratio $BR(\hbb)$ as a
function of $\Mh$ and $\MA$. For values of $\MA \gsim 250 \gev$
there is a non-negligible difference between \onel\ and \twol\ order,
where at the \twol\ level the branching ratio is slightly enhanced. 
Compared in terms of $\Mh$ there is nearly no change for small values
of $\Mh$ in the low and in the high $\tb$ case. 
Here $BR(\hbb)$ is changed by less than about 1\%, see \reffi{fig:BRhbb}. 
$BR(\htautau)$ is
increased by less than about 2\%.
$BR(\hcc)$ can be increased at the
\twol\ level by ${\cal O}(50\%)$, but remains numerically relatively
small. For $\tb = 40$ the
main difference arises at the endpoints of the spectrum, again due to
the fact that different Higgs boson masses can be obtained at the
\onel\ and at the \twol\ level. For $\tb = 3$, however, also several
GeV below the kinematical endpoints there is a sizable effect on
$BR(\hbb)$. 
Thus, in the experimentally allowed region of $\Mh$, the \twol\
corrections can have an important effect on $BR(\hbb)$.

\smallskip
Higgs boson search, especially at \epem\ colliders, often relies on $b$
search, since on one hand the lightest $\cp$-even 
Higgs boson decays dominantly into $b\bar{b}$ and on the other hand
$b$~tagging can be performed with high efficiency. For some
combinations of parameters, however, $\Ga(\hbb)$ can become very small
and thus $BR(\hbb)$ can approach zero as a consequence of large
Higgs-boson propagator corrections or large gluino vertex-corrections,
making Higgs boson search 
possibly very difficult for these parameters. 
Within the effective potential approach this kind of effect has first been
observed in \citere{loinazwells}, recent analyses 
investigating the parameter regions where $BR(\hbb)$ is suppressed
can be found in \citeres{gluinoresum1,higgssearch2}.
In order to have reliable
predictions for these regions of parameter space a full calculation of
the \onel\ vertex
corrections, including all $\oa$ contributions, would be necessary. Here
we demonstrate the effect of the \twol\ propagator corrections on
the values of the parameters, especially of $\MA$, for which
$BR(\hbb)$ goes to zero. We 
also show the impact of the inclusion of the momentum dependence of
the Higgs boson self-energies (see \refeq{higgsmassmatrixnondiag} and
(\ref{zeroexternalmomentum})), that is often neglected in 
phenomenological analyses of the decays of the lightest $\cp$-even
Higgs boson.

\begin{figure}[ht!]
\begin{center}
\mbox{
\psfig{figure=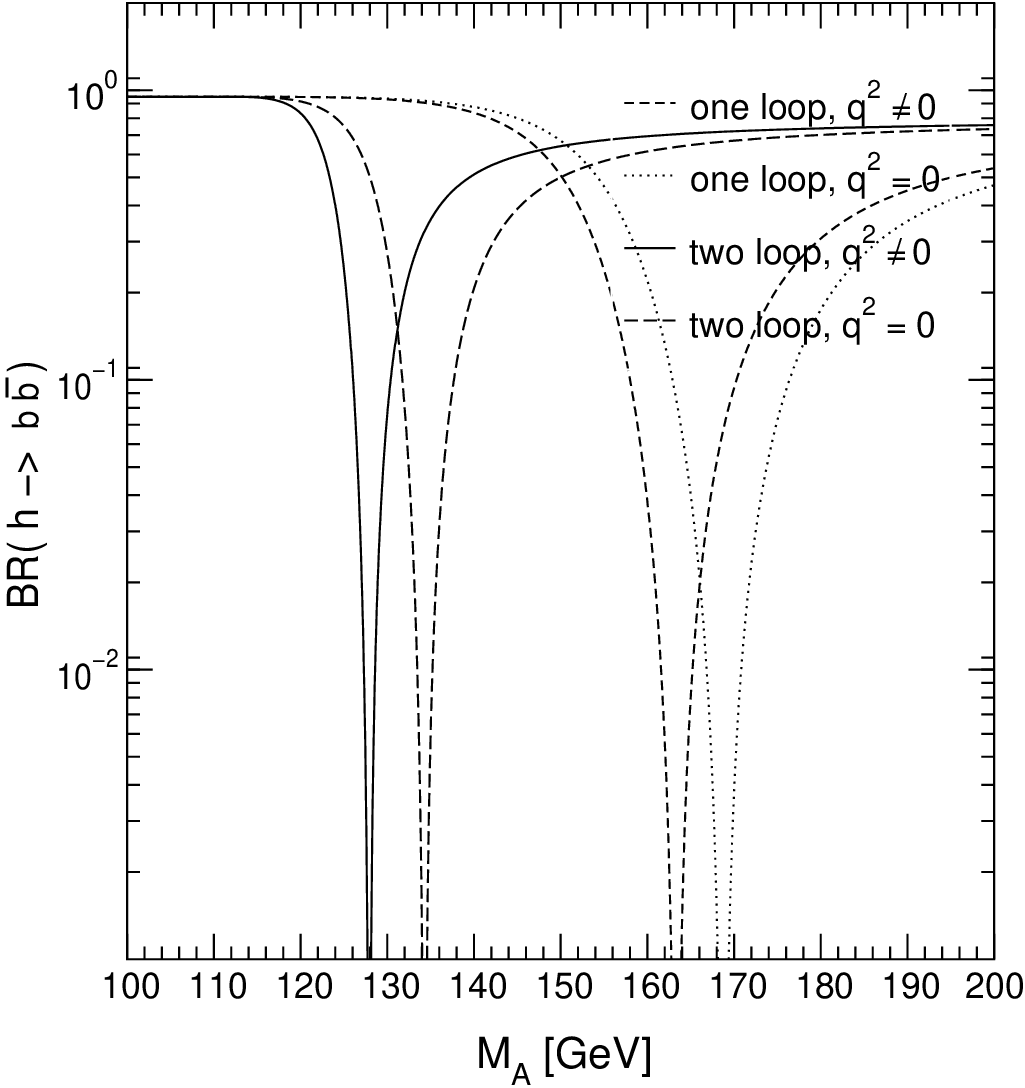,width=10cm,height=8cm}}
\end{center}
\caption[]{{\it
$BR(\hbb)$ is shown as a function
of $\MA$. The Higgs boson self-energies are evaluated at the \onel\
and at the \twol\ level with and without momentum dependence (see
\refeq{zeroexternalmomentum}). The other parameters are
$\tb = 25$, $\msq = 500 \gev$, $\mgl = 400 \gev$, $M_2 = 400 \gev$,
$\Xt = 400 \gev$, $\Ab = \At$, $\mu = -1000 \gev$. 
}}
\label{fig:brhbbzero}
\end{figure}

In \reffi{fig:brhbbzero} $BR(\hbb)$ is shown as a function
of $\MA$. The Higgs boson self-energies are evaluated at the \onel\
and at the \twol\ level with and without momentum dependence (see
\refeq{zeroexternalmomentum}). The other parameters are
$\tb = 25$, $\msq = 500 \gev$, $\mgl = 400 \gev$, $M_2 = 400 \gev$,
$\Xt = 400 \gev$, $\Ab = \At$, $\mu = -1000 \gev$. The inclusion of
the \twol\ propagator corrections shifts the $\MA$ value for which 
$BR(\hbb)$ becomes very small by about $-35 \gev$. The
inclusion of the momentum dependence of the Higgs boson self-energies
induces another shift of about $-6 \gev$. In order to have reliable
phenomenological predictions for the problematic $\MA$ values the
\twol\ corrections as well as the inclusion of the momentum dependence
is necessary. 
Note that the inclusion of the gluino vertex corrections as well as
the purely weak vertex corrections can also have a large impact on the
critical $\MA$ values.


\subsection{Effects of the gluino vertex corrections}
\label{subsec:gluinoeffect}

In this subsection we present the effect of the gluino-exchange
contribution to the $hf\bar{f}$ vertex corrections. 
These corrections have been neglected so far in most 
phenomenological analyses%
\footnote{
The gluino-exchange contributions are currently
incorporated into HDECAY~\cite{hdecay,spirix}. Concerning the
discovery potential of LEP, the Tevatron and the LHC, the gluino corrections
have also been studied recently in \citeres{higgssearch2,gluinoresum2}.
}%
.

\begin{figure}[ht!]
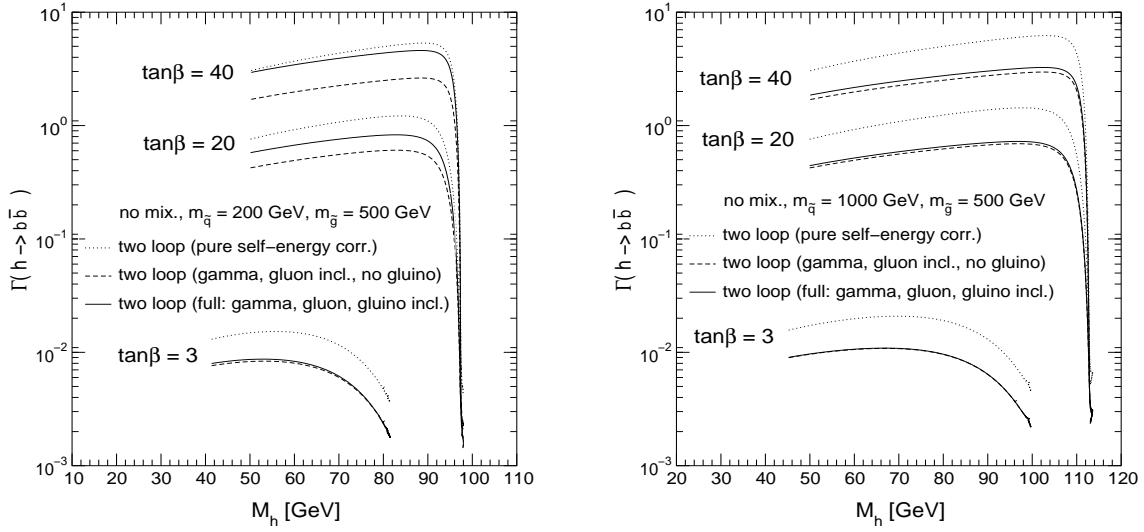

\begin{center}
\mbox{
\psfig{figure=gh2lC6.bw.eps,width=7cm,height=7cm}}
\hspace{1.5em}
\mbox{
\psfig{figure=gh2lC7.bw.eps,width=7cm,height=7cm}}
\end{center}
\caption[]{{\it
$\Ga(\hbb)$ is shown as a function
of $\Mh$ for three values of $\tb$. The Higgs-propagator corrections have 
been evaluated at the \twol\ level in the no-mixing scenario.
The dotted curves shows the results containing only the pure 
self-energy corrections. The results given in the 
dashed curves in addition contain the QED correction and the gluon-exchange
contribution. The solid curves show the full result, including
also the gluino correction. The other parameters are
$\mu = -100 \gev$, $M_2 = \msq$, $\Ab = \At$.
}}
\label{fig:Ghbbgluino}
\end{figure}

\begin{figure}[ht!]
\begin{center}
\mbox{
\psfig{figure=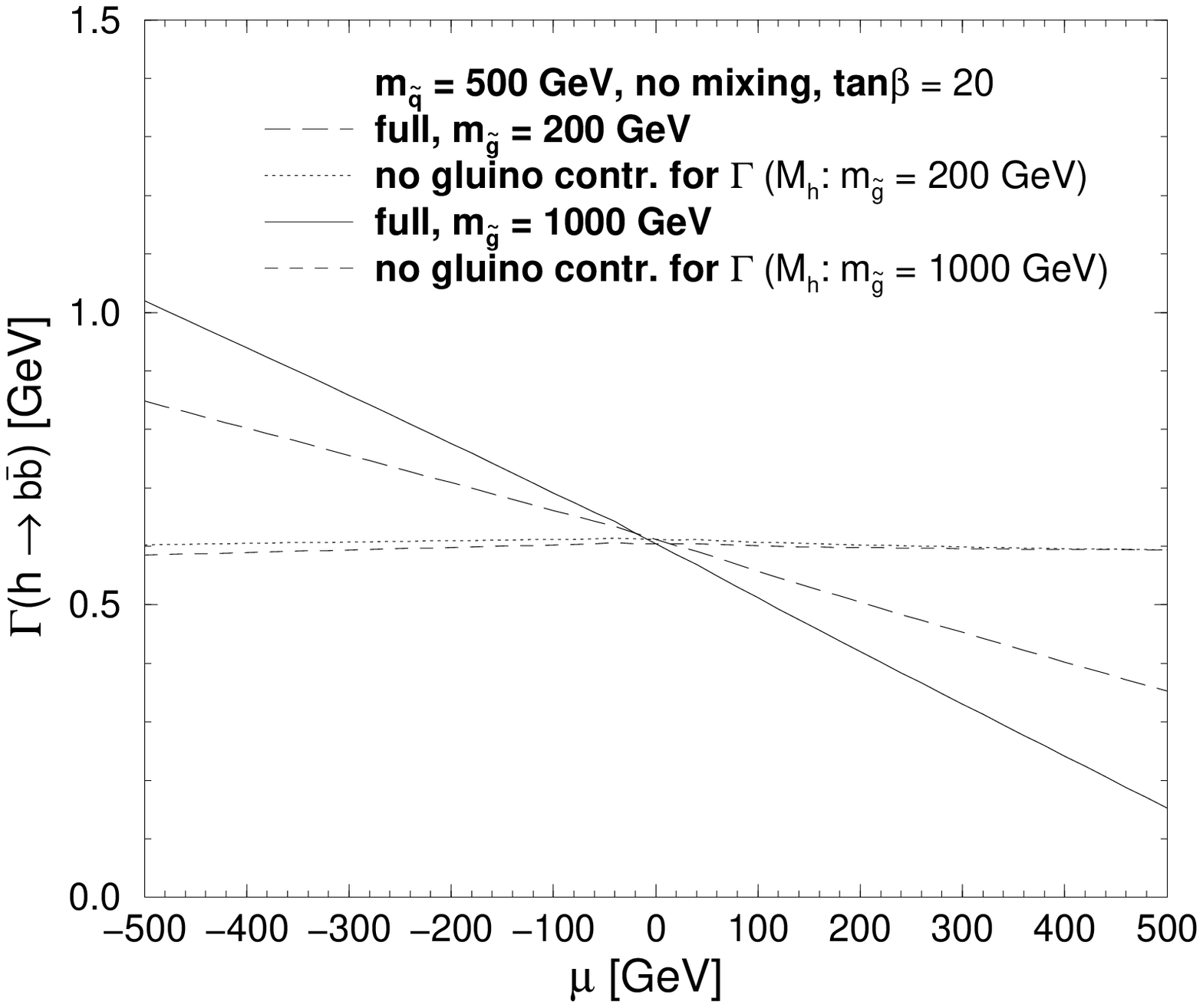,width=7cm,height=7cm}}
\hspace{1.5em}
\mbox{
\psfig{figure=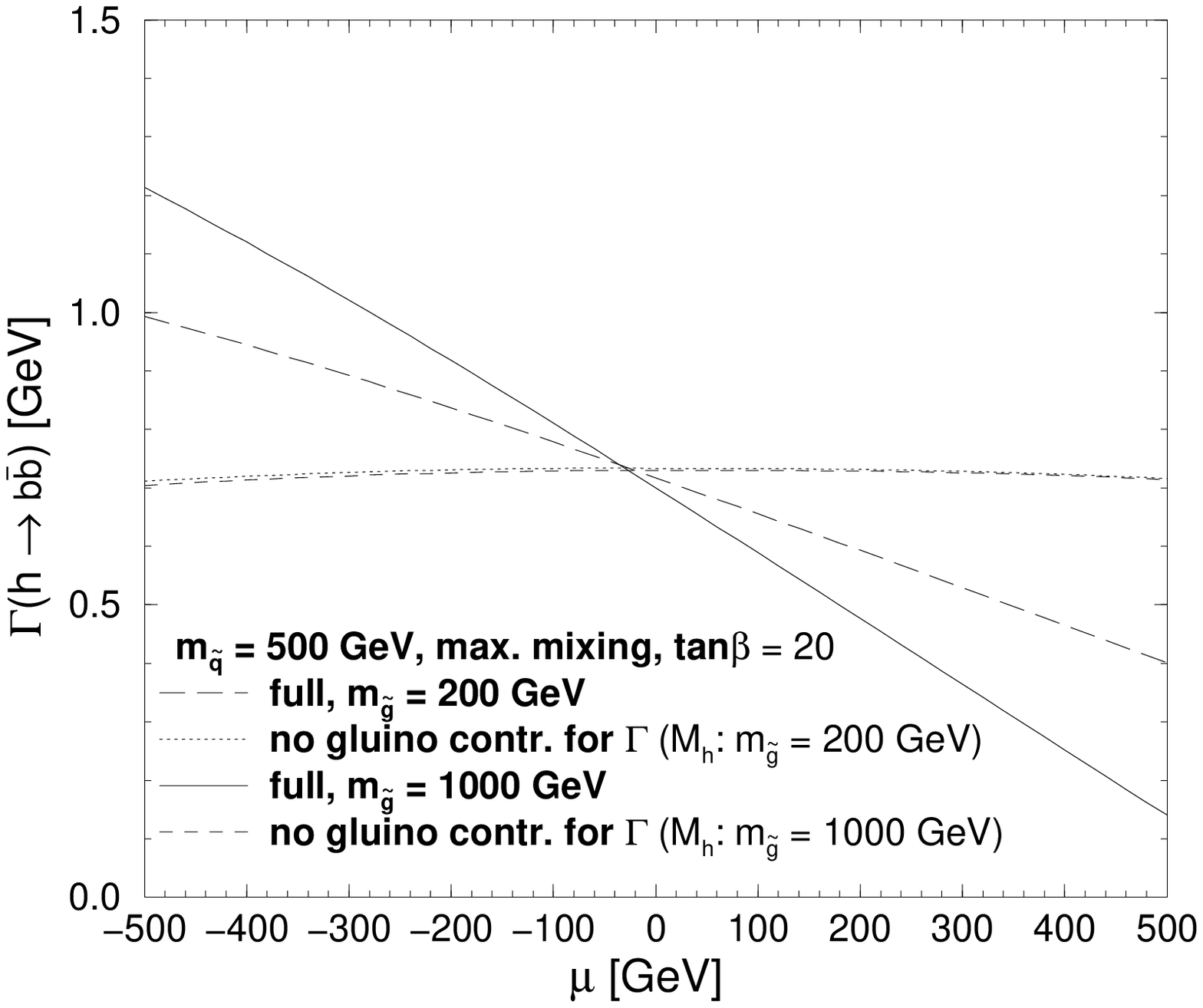,width=7cm,height=7cm}}
\end{center}
\caption[]{{\it
$\Ga(\hbb)$ is shown as a function
of $\mu$ for $\Tb = 20$ and two different values of $\mgl$ in the
no-mixing and the maximal-mixing scenario.
The other parameters are
$\msq = 500 \gev$, $M_2 = 500 \gev$, $\Ab = \At$, $\MA = 100 \gev$.
}}
\label{fig:Ghbbgluinomue}
\end{figure}

\reffi{fig:Ghbbgluino} shows $\Ga(\hbb)$ in three steps of accuracy:
the dotted curves contain only the pure self-energy correction, the
dashed curves contain in addition the QED and the gluon-exchange
correction. The solid curves show the full results, including also the
gluino-exchange correction. The results are shown for the no mixing
scenario, $\mu = -100 \gev$, 
$\mgl = 500 \gev$, $\msq = 200, 1000 \gev$ in the left and right part
of \reffi{fig:Ghbbgluino}, respectively. For $\tb$, three values have
been chosen: $\tb = 3, 20, 40$.

The left plot of \reffi{fig:Ghbbgluino} corresponds to a small soft
SUSY-breaking scale, $\msq = 200 \gev$. The effect of the gluon
contribution is large and negative, the effect of the
QED correction is small. For this combination of $\mgl$, $\msq$
and $\mu$ the effect of the gluino correction is large and positive as
can be seen from the transition from the dashed to the solid
curves. For $\Tb = 40$ it nearly compensates the gluon effect, for 
$\tb = 20$ it amounts up to 20\% of the gluonic correction, while for 
$\tb = 3$ the gluino-exchange contribution is negligible. Note that we
have chosen a relatively small value of $\mu$, $\mu = -100 \gev$. For
larger values of $|\mu|$ even larger correction can be obtained. Hence
neglecting the gluino-exchange correction in the large $\tb$
scenario can lead to results which
deviate by 50\% from the full $\oas$ calculation (see also
\refse{subsubsec:decaywidthgluino}). 
The right plot of \reffi{fig:Ghbbgluino} corresponds to $\msq = 1000 \gev$. 
The gluino-exchange effects are still visible, but much smaller
than for $\msq = 200 \gev$. The same observation has already been
made in \citere{hff1l}.
In the maximal-mixing scenario we find qualitatively the same behavior
for the gluino-exchange corrections as in the no-mixing scenario.

In \reffi{fig:Ghbbgluinomue} the pure gluino-exchange effect is
shown as a function of $\mu$. This effect increases with rising%
\footnote{
This is correct for all values of $\mgl$ considered in this work. A
maximal effect is reached around $\mgl \approx 1500 \gev$.
The decoupling of the gluino takes place only for very large values,
$\mgl \gsim 5000 \gev$.
}
$\mgl$ and $|\mu|$, where for
negative (positive) $\mu$ there is an enhancement (a decrease) in
$\Ga(\hbb)$. The size of the gluino-exchange contribution also depends
on $\MA$, where larger effects correspond to smaller values of $\MA$,
see also \citere{hff1l}. The small difference between the curves where the
decay rate has been calculated without gluino contribution is due to
the variation of $\Mh$ induced by different values of $\mgl$ which
enters at $\oaas$.
\reffi{fig:Ghbbgluinomue} demonstrates again
that neglecting the gluino contribution in the fermion decay rates can
yield (strongly) misleading results.

\begin{figure}[ht!]
\begin{center}
\mbox{
\psfig{figure=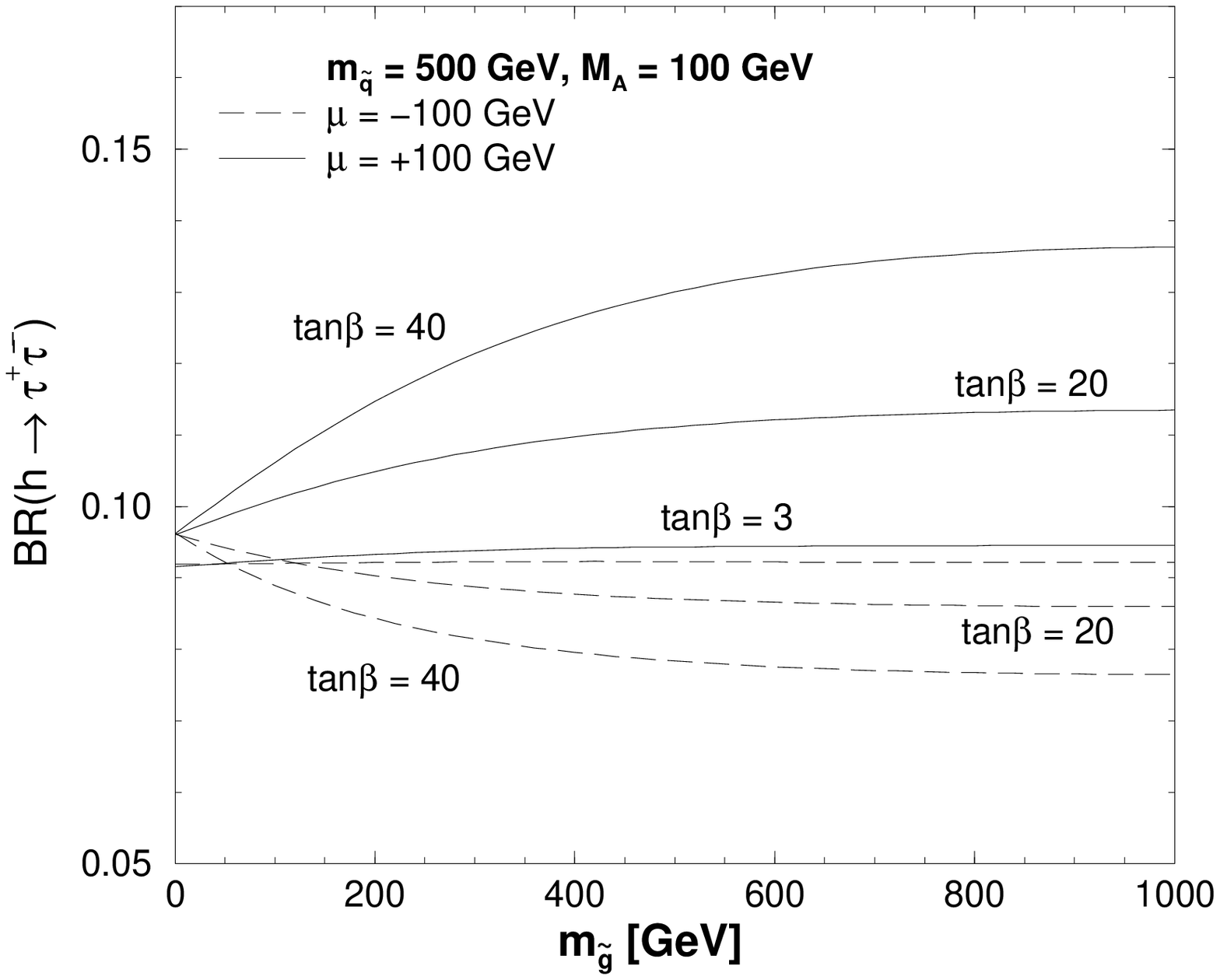,width=7cm,height=6.5cm}}
\hspace{1.5em}
\mbox{
\psfig{figure=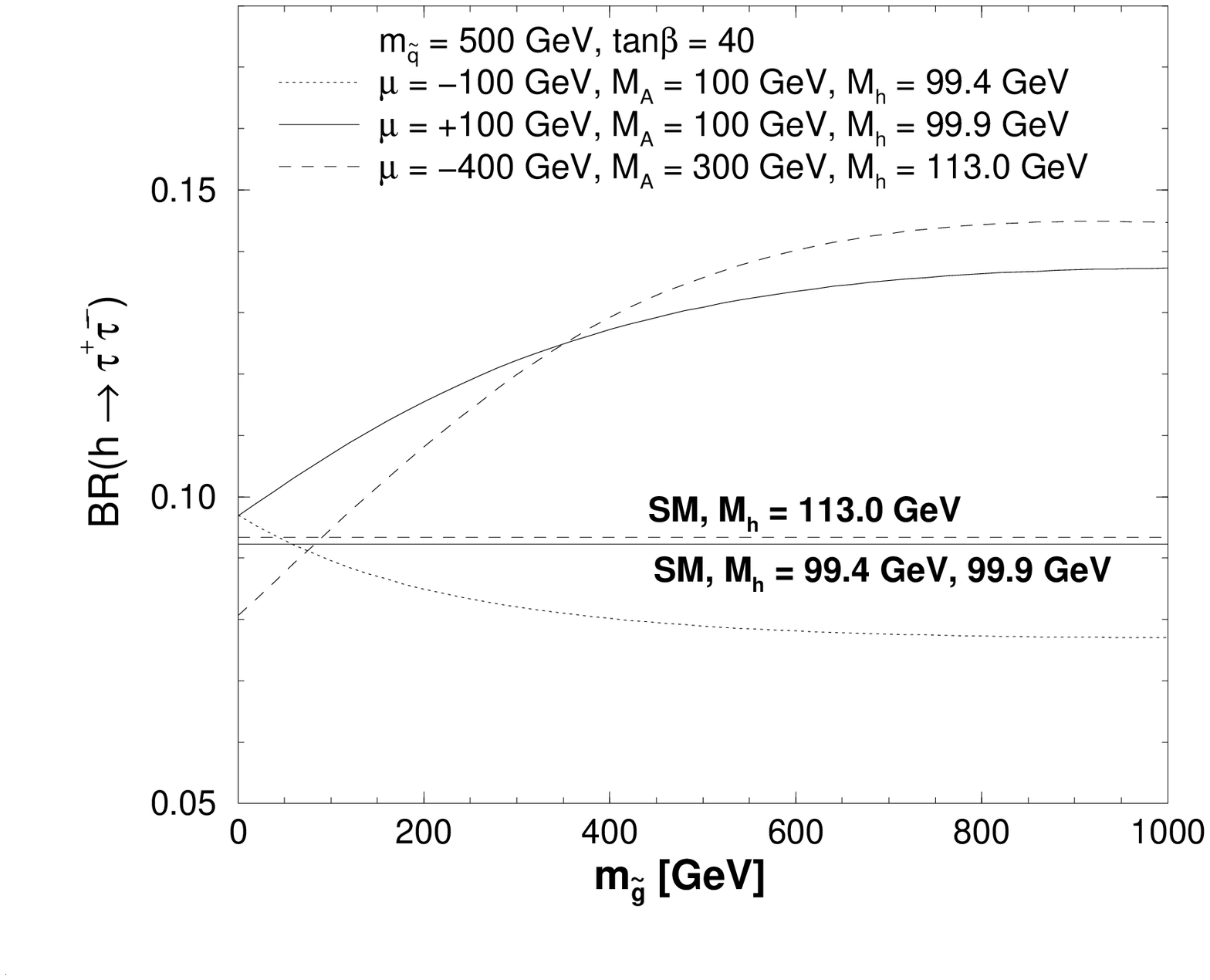,width=7cm,height=6.5cm}}
\end{center}
\caption[]{{\it
$BR(\htautau)$ is shown as a function of $\mgl$.
The left plot corresponds to three different values of 
$\tb$ and $\mu = \pm 100 \gev$.
The right plot corresponds to $\Tb = 40$ and the two scenarios 
$\MA = 100 \gev, \mu = \pm 100 \gev$ and 
$\MA = 300 \gev, \mu = -400 \gev$. 
The other parameters in both plots are
$\msq = 500 \gev, \Xt = 500 \gev, \MA = 100 \gev$, $M_2 = 500 \gev$, 
$\Ab = \At$.
The Higgs boson masses given in the legend are averaged masses over
the interval $0 < \mgl < 1000 \gev$ where the variation of $\Mh$ is
about $\pm 1 \gev$ for the above considered parameters.
The horizontal lines represent the SM result using the corresponding
Higgs boson masses.
}}
\label{fig:BRttgluino}
\end{figure}

\bigskip
The gluino exchange contribution has only a relatively small impact on  
$BR(\hbb)$. It can have a large influence, on the other hand, on
$BR(\htautau)$. Both branching ratios
are expected to be measurable at the same level of accuracy, see
e.g. \citere{teslacdr}. 
While the Higgs-propagator contributions are universal corrections
that affect $\Ga(\hbb)$ and $\Ga(\htautau)$ in the same way (i.e.\ the
influence on the effective coupling is the same in both cases), the
gluino corrections, 
which influence only $\Ga(\hbb)$, can lead to a different behavior of
the two decay widths.
In \reffi{fig:BRttgluino} we show $BR(\htautau)$ as a function of
$\mgl$.
The left plot corresponds to three different values of 
$\tb$ and $\mu = \pm 100 \gev$, $\MA = 100 \gev$. 
Here we have furthermore chosen $\msq = 500 \gev$ and moderate mixing,
i.e.\ $\Xt = \msq$; we find similar results for the other mixing scenarios.
For positive (negative) $\mu$ the decay rate 
$\Ga(\hbb)$ is reduced (enhanced), see \reffi{fig:Ghbbgluinomue}, 
thus $BR(\htautau)$ is increased (decreased) by up to 50\%. 
For large values of $\mgl$ and $\tb$, $BR(\htautau)$ can thus be considerably
different from the case where the gluino-exchange contribution to
$\Ga(\hbb)$ has been neglected.
This becomes even more apparent in the right plot of
\reffi{fig:BRttgluino}, where the MSSM result, including gluino exchange
contribution, is compared to the SM result. 
Here we also show the scenario with 
$\MA = 300 \gev$ and $\mu = - 400 \gev$, 
where the Higgs sector of the MSSM behaves SM like (i.e.\ the lightest
Higgs boson has almost SM couplings, all other Higgs bosons are heavy).
The horizontal lines represent the SM values for the respective
Higgs boson masses. The two Higgs boson masses for each line give
similar results so that the lines are indistinguishable in the
plot. These masses correspond to an averaged value obtained in the MSSM 
in the interval $0 < \mgl < 1000 \gev$, where the variation of $\Mh$
is about $\pm 1 \gev$ (with our choice of $\msq$ and $\Xt$). 
Since the gluino decouples very slowly, there is no
decoupling effect for $\mgl \lsim 1000 \gev$ and
the MSSM results can be considerably different from
the SM result, for $\MA = 100 \gev$ as well as for $\MA = 300 \gev$,
i.e.\ even where the MSSM Higgs sector behaves otherwise SM like.
The deviation can amount up to 50\% for $\MA =$ \order{100 \gev} and up
to 30\% for $\MA =$ \order{300 \gev}. Thus the measurement of
$BR(\htautau)$ can provide a distinction between the SM and the MSSM
even for relatively large $\MA$ and large $\mgl$ if also $|\mu|$ and
$\tb$ are sufficiently large. The branching ratio
$BR(\hbb)$ on the other hand, changes only by a few per cent for these
parameters, so that this change would be much harder to measure.

\subsection{The $\aeff$-approximation}
\label{subsec:aeffeffect}

\begin{figure}[ht!]
\begin{center}
\mbox{
\psfig{figure=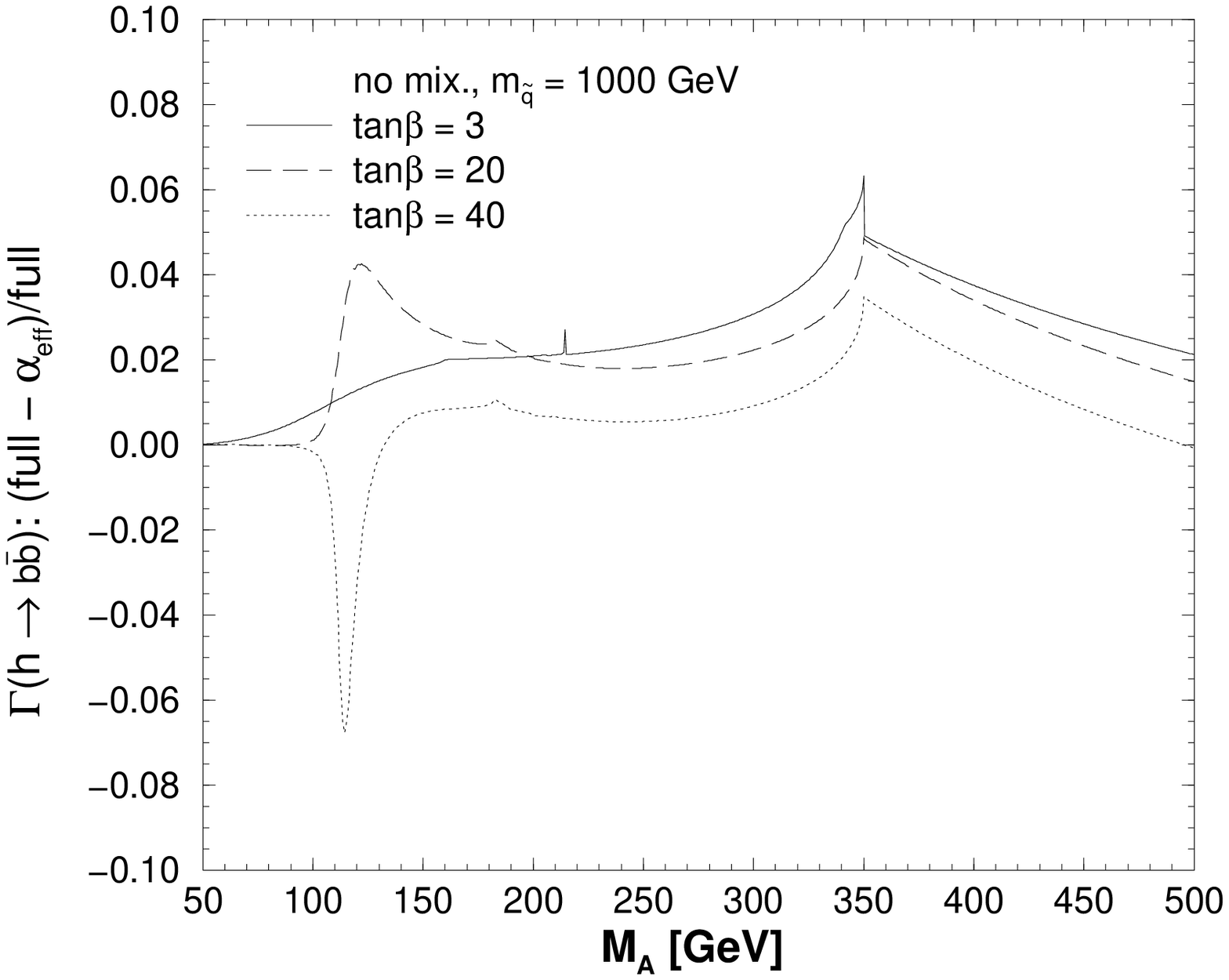,width=7cm,height=6.5cm}}
\hspace{1.5em}
\mbox{
\psfig{figure=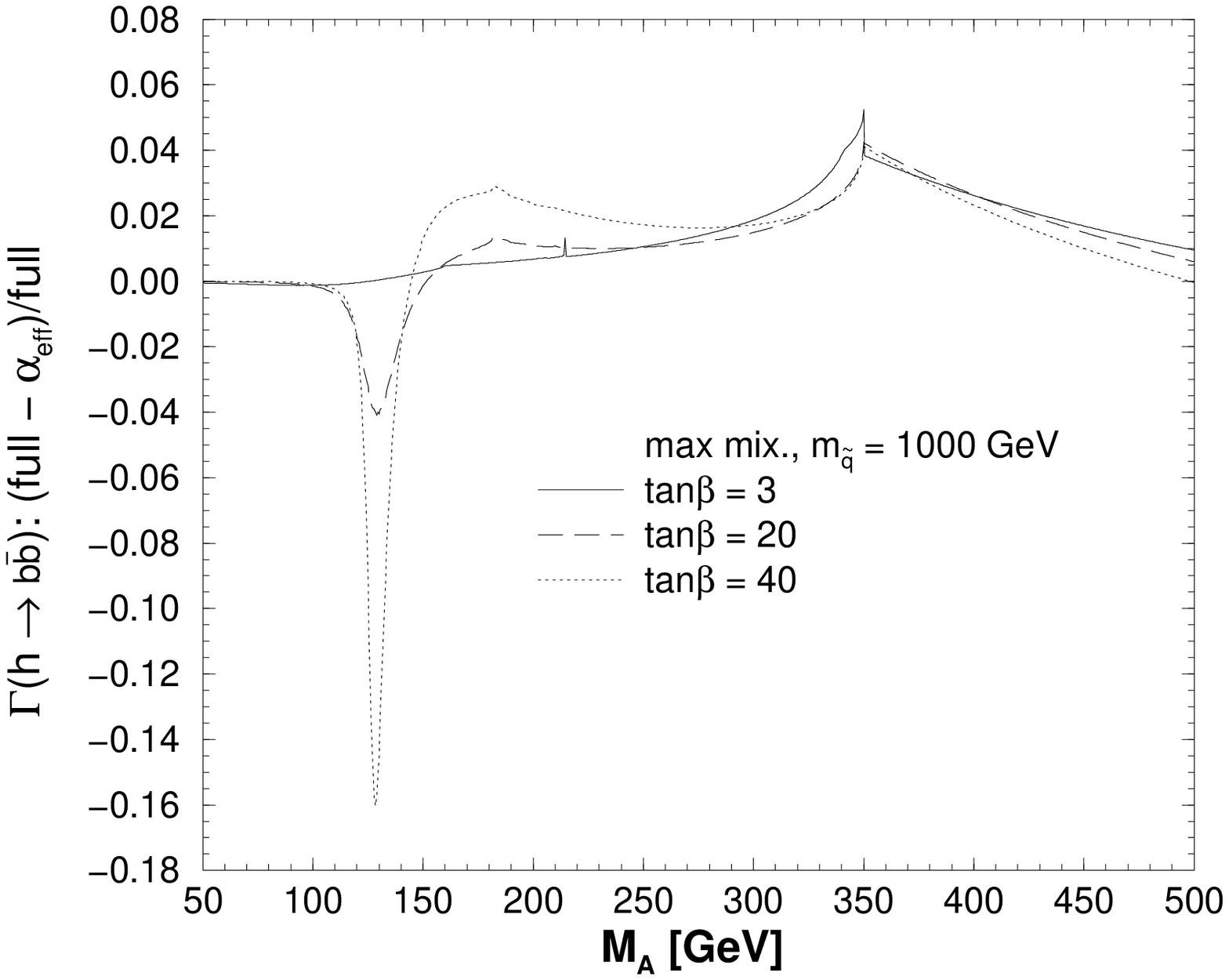,width=7cm,height=6.5cm}}
\end{center}
\caption[]{{\it
$\De\Ga(\hbb) = 
(\Ga^{\rm full}(\hbb) - \Ga^{\aeff}(\hbb))/\Ga^{\rm full}(\hbb)$ 
is shown as a function of $\MA$ for three values of $\tb$.
The QED, gluon- and gluino-contributions are neglected here. 
The other parameters are 
$\mu = -100 \gev$, $M_2 = \msq$, $\mgl = 500 \gev$, $\Ab = \At$, 
$\Tb = 3, 20, 40$. The results are given in the minimal- and the
maximal-mixing scenario.
}}
\label{fig:RCGhbbaeff}
\end{figure}
%
\begin{figure}[ht!]
\begin{center}
\mbox{
\psfig{figure=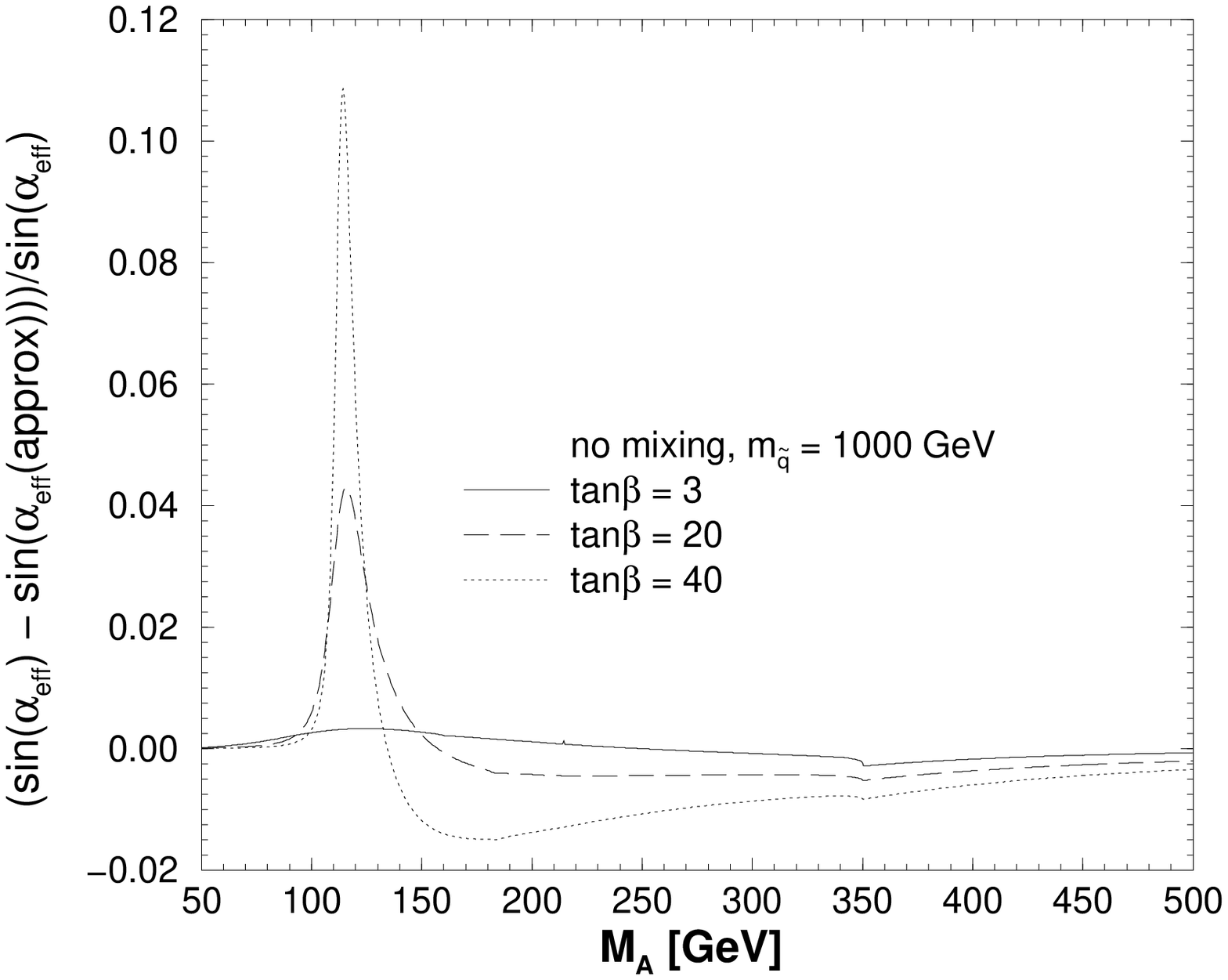,width=7cm,height=6.5cm}}
\hspace{1.5em}
\mbox{
\psfig{figure=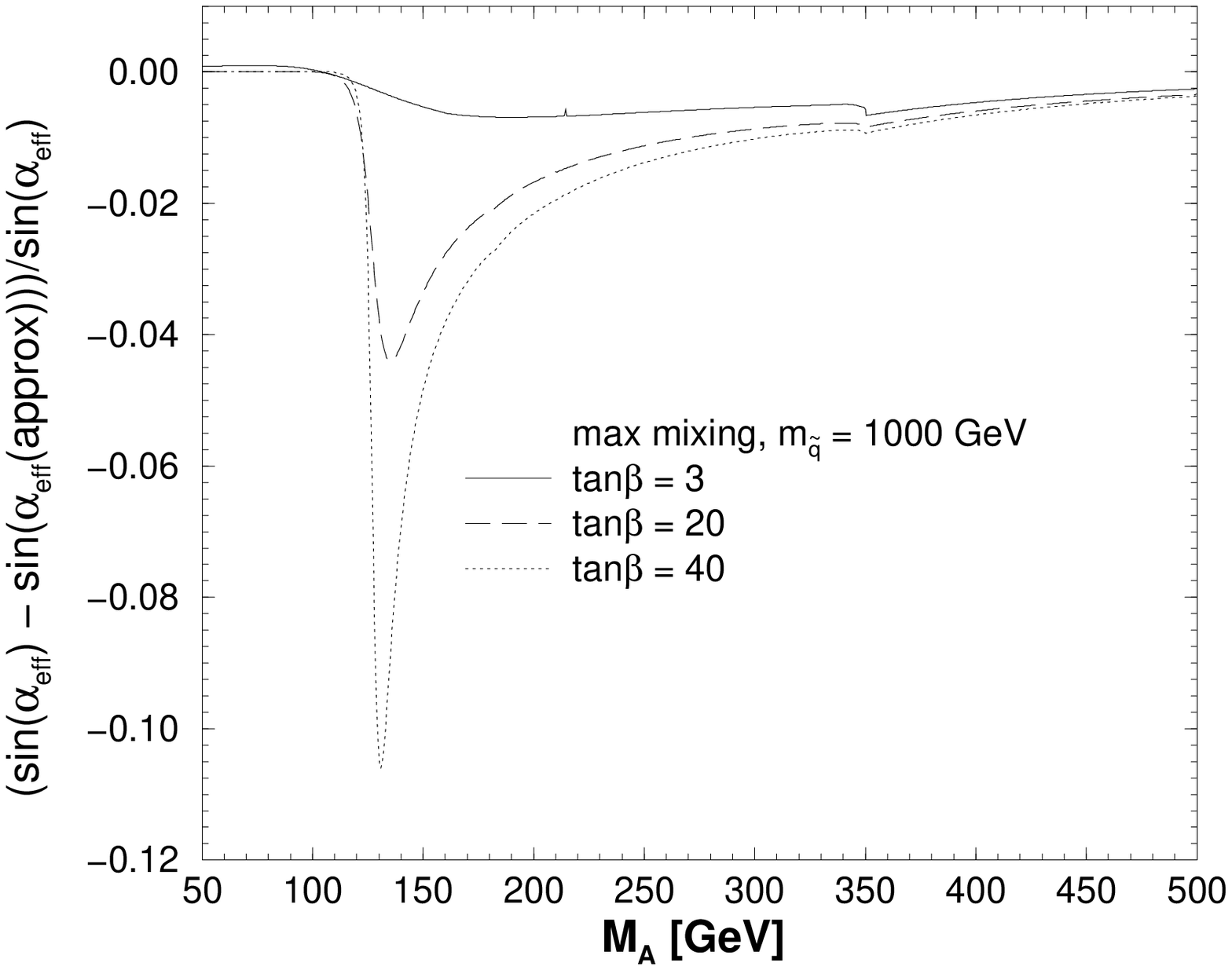,width=7cm,height=6.5cm}}
\end{center}
\caption[]{{\it
The relative difference $(\sin\aeff - \sin\aeffapprox)/\sin\aeff$  
(see \refeqs{hseq2zero} and~(\ref{hselle})) is shown as a function
of $\MA$ for three values of $\tb$ in the no mixing and the maximal
mixing scenario. The other parameters are chosen as in
\reffi{fig:RCGhbbaeff}. 
}}
\label{fig:RCsaeff}
\end{figure}

In this section we investigate the quality of the
$\aeff$-approximation. In \reffi{fig:RCGhbbaeff} we display the
relative difference between the full result~(\ref{hsefull}) and the
$\aeff$ result, where the external momentum of the Higgs self-energies
has been neglected, see \refeq{hseq2zero}.
The relative difference $\De\Ga(\hbb) = 
(\Ga^{\rm full}(\hbb) - \Ga^{\aeff}(\hbb))/\Ga^{\rm full}(\hbb)$ is
shown as a function of $\MA$ for $\msq = 1000 \gev$ and for three values
of $\tb$ in the no mixing and the maximal mixing scenario. 
Large deviations
occur only in the region $100 \gev \lsim \MA \lsim 150 \gev$, especially
for large $\tb$. In this region of parameter space the values of $\Mh$
and $\MH$ are very close to each other. This results in a high
sensitivity to small deviations in the Higgs boson self-energies
entering the Higgs-boson mass
matrix~(\ref{higgsmassmatrixnondiag}),~(\ref{deltaalpha}). 
Another source of differences between the full and the approximate
calculation is the threshold $\MA = 2\,\mt = 350 \gev$ in the \onel\
contribution, originating from the top-loop diagram in the $A$
self-energy and in the $AZ$
self-energy (see \citere{mhiggsf1l}). Here the deviation can amount up to 6\%.

In \reffi{fig:RCsaeff} we compare the $\aeff$ result (\ref{hseq2zero})
with the $\aeffapprox$ result (\ref{hselle}), where the
Higgs boson self-energies have been approximated by the compact
analytical expression obtained in~\citere{mhiggslle}.
\reffi{fig:RCsaeff}
displays  the relative difference in the effective mixing angles,
$(\sin\aeff - \sin\aeffapprox)/\sin\aeff$. Via \refeq{ampeffhbb} 
$\Saeff$ directly determines 
the decay width $\Ga(\hbb)$. The result is shown for 
$\msq = 1000 \gev$, for three values of $\tb$ in the minimal and the
maximal mixing scenario.
Apart from the region around
$\MA \approx 120 \gev$ (compare \reffi{fig:RCGhbbaeff}) both effective
angles agree better than 3\% with each other. 

Concerning the comparison of the $\aeff$-approximations in terms of
$\Mh$ (which is not plotted here), 
due to the neglected external momentum or the neglected subdominant
one- or \twol\ terms, $\Mh$ receives a slight shift.
Besides this kinematical effect, 
the decay rate is approximated rather well for most of the $\Mh$ values: 
independently of $\msq$, the differences stay mostly below 2-4\%, 
for the no-mixing case as well as for the maximal-mixing case.
Only at the endpoints of the spectrum, due to the different
Higgs-boson mass values, the difference is not negligible.


\section{Comparison with the renormalization group approach}
\label{sec:rgcomparison}

\begin{figure}[ht!]
\begin{center}
\mbox{
\psfig{figure=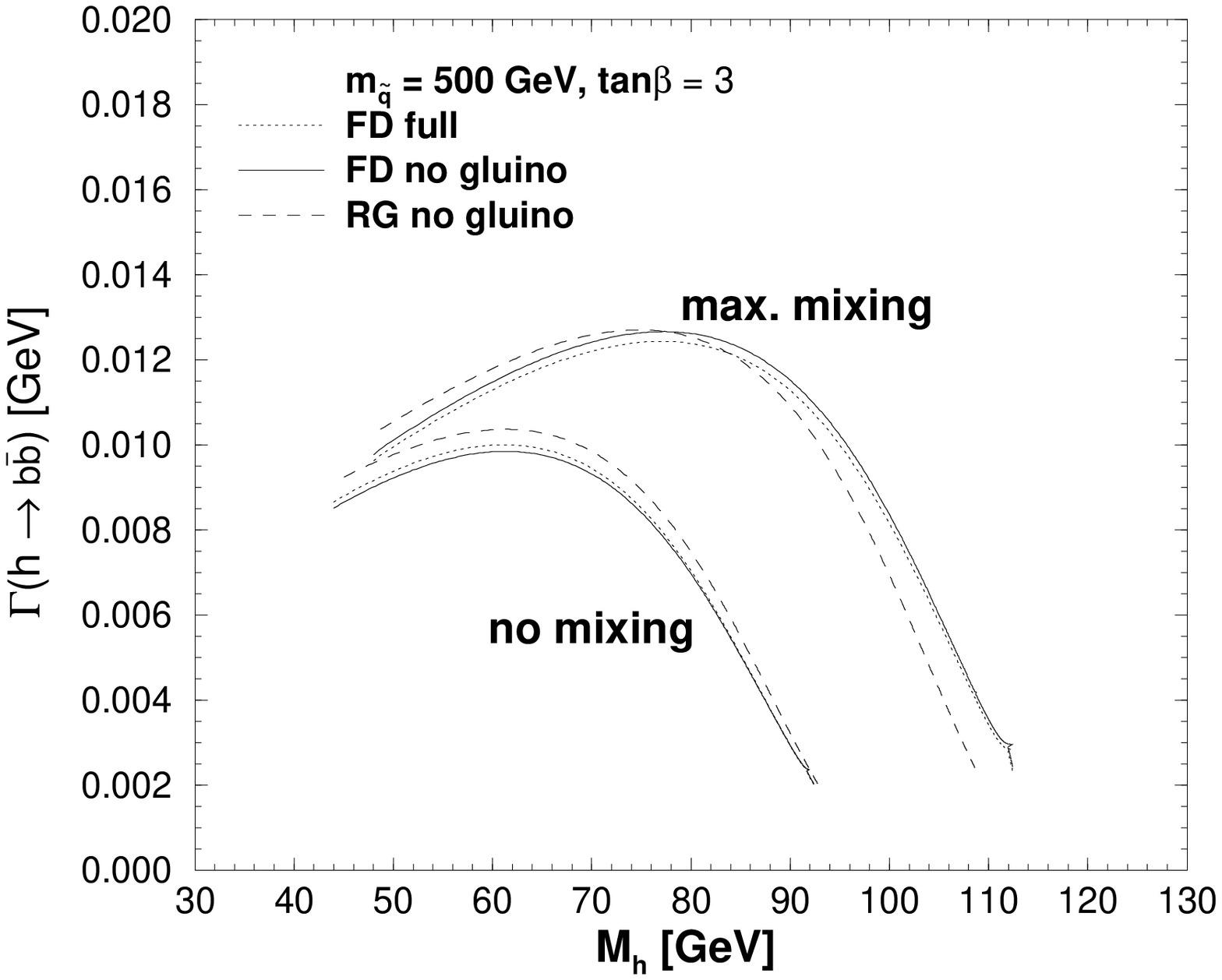,width=7cm,height=5.4cm}}
\hspace{1.5em}
\mbox{
\psfig{figure=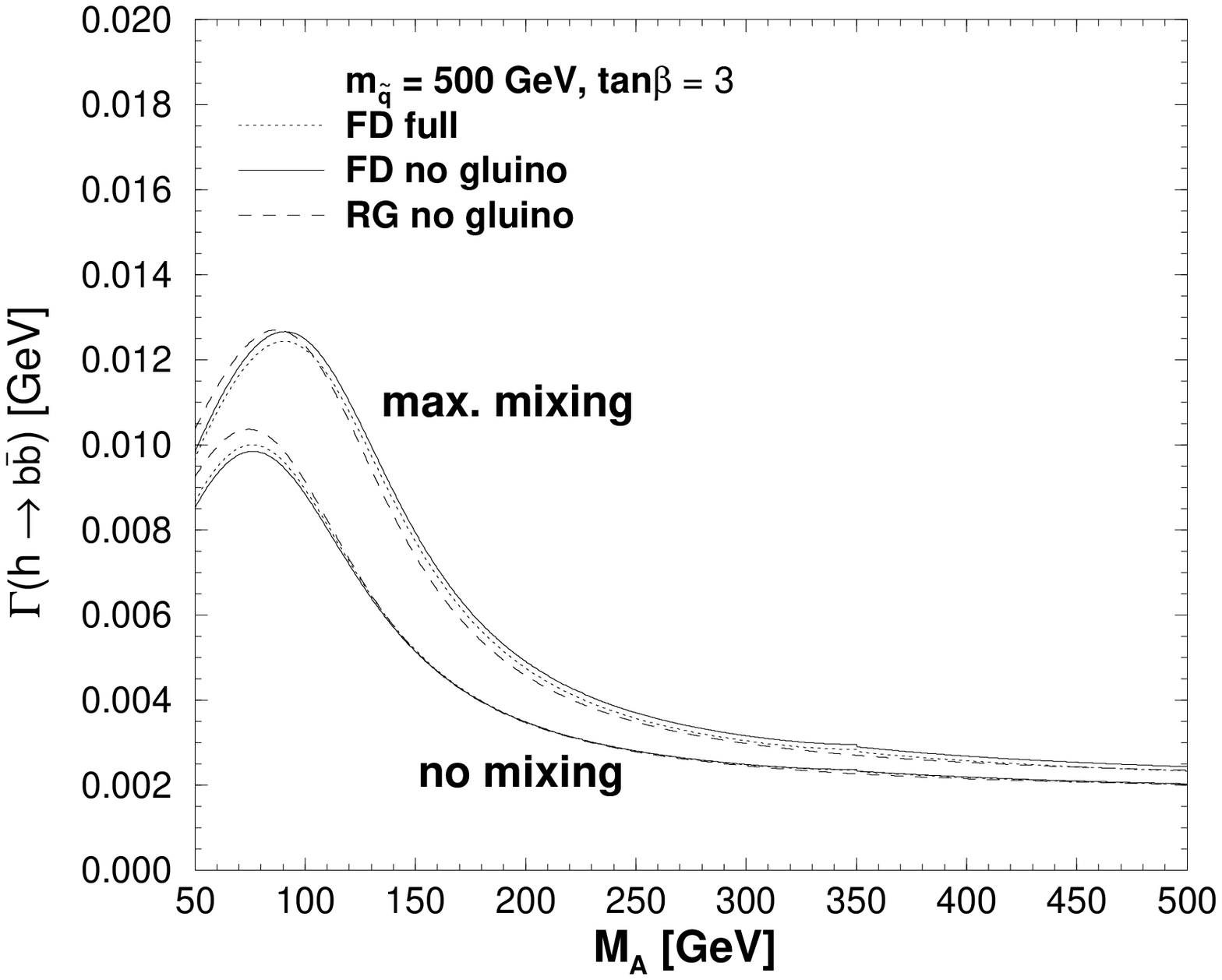,width=7cm,height=5.4cm}}
\end{center}
\begin{center}
\mbox{
\psfig{figure=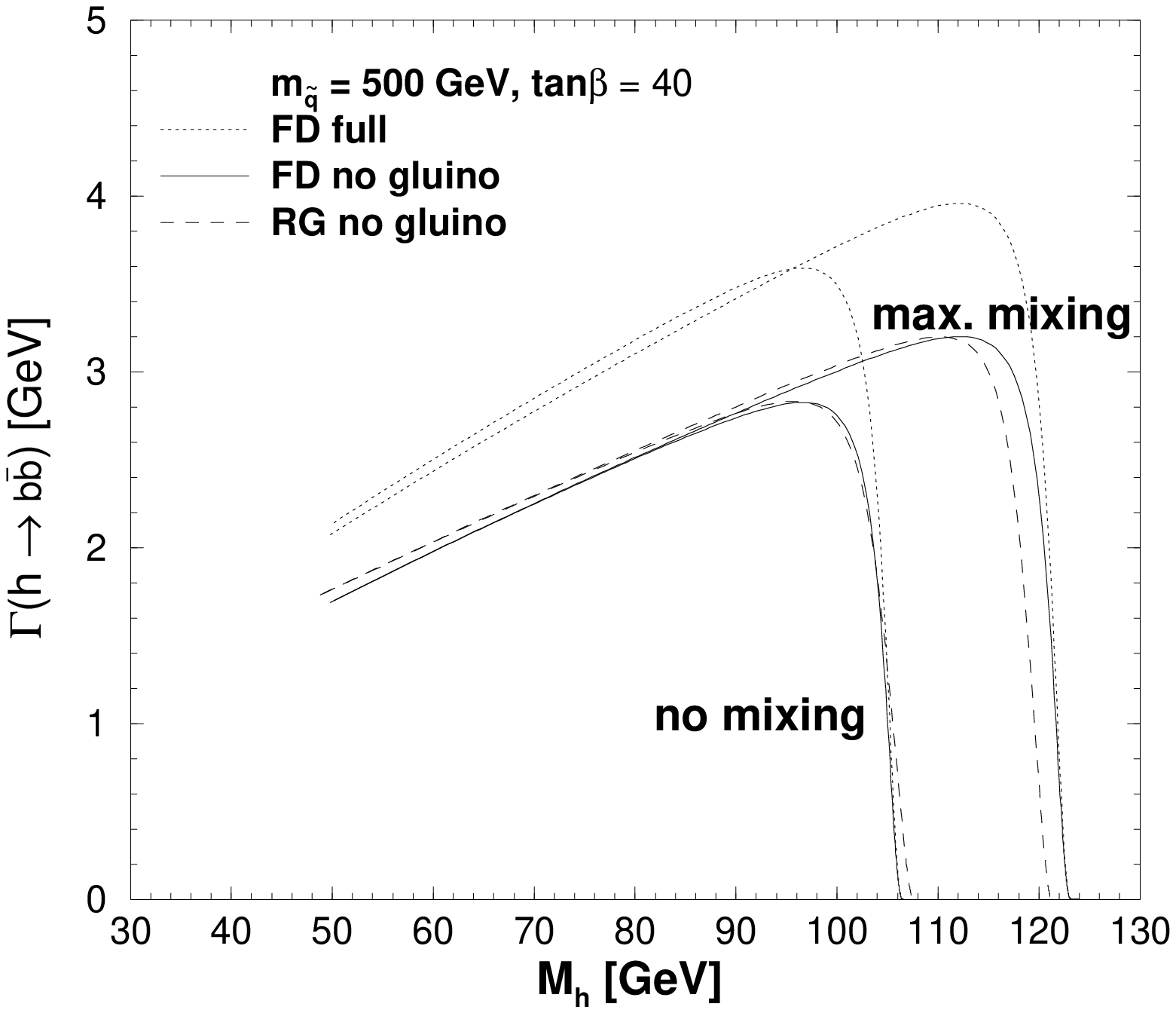,width=7cm,height=5.4cm}}
\hspace{1.5em}
\mbox{
\psfig{figure=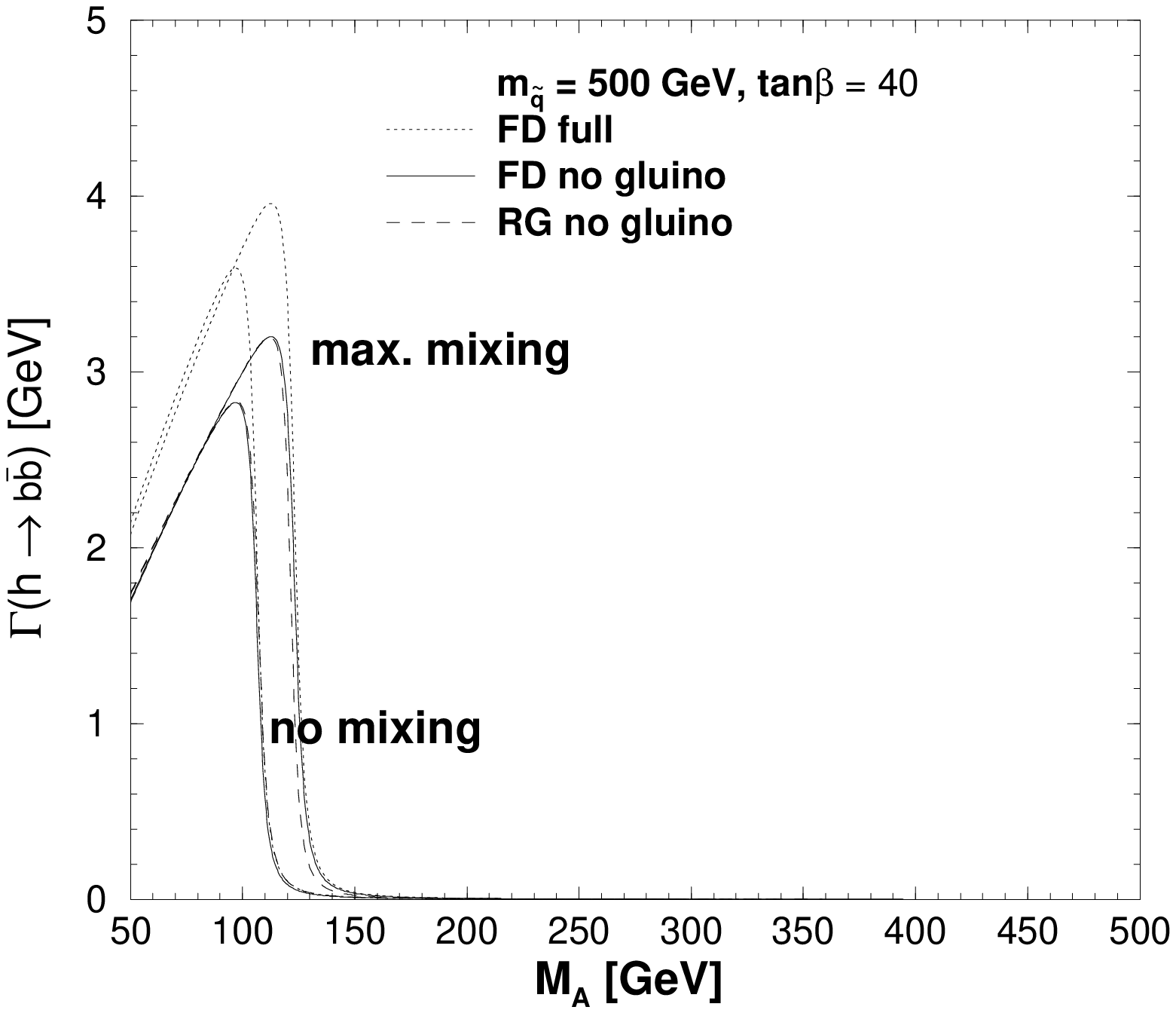,width=7cm,height=5.4cm}}
\end{center}
\begin{center}
\mbox{
\psfig{figure=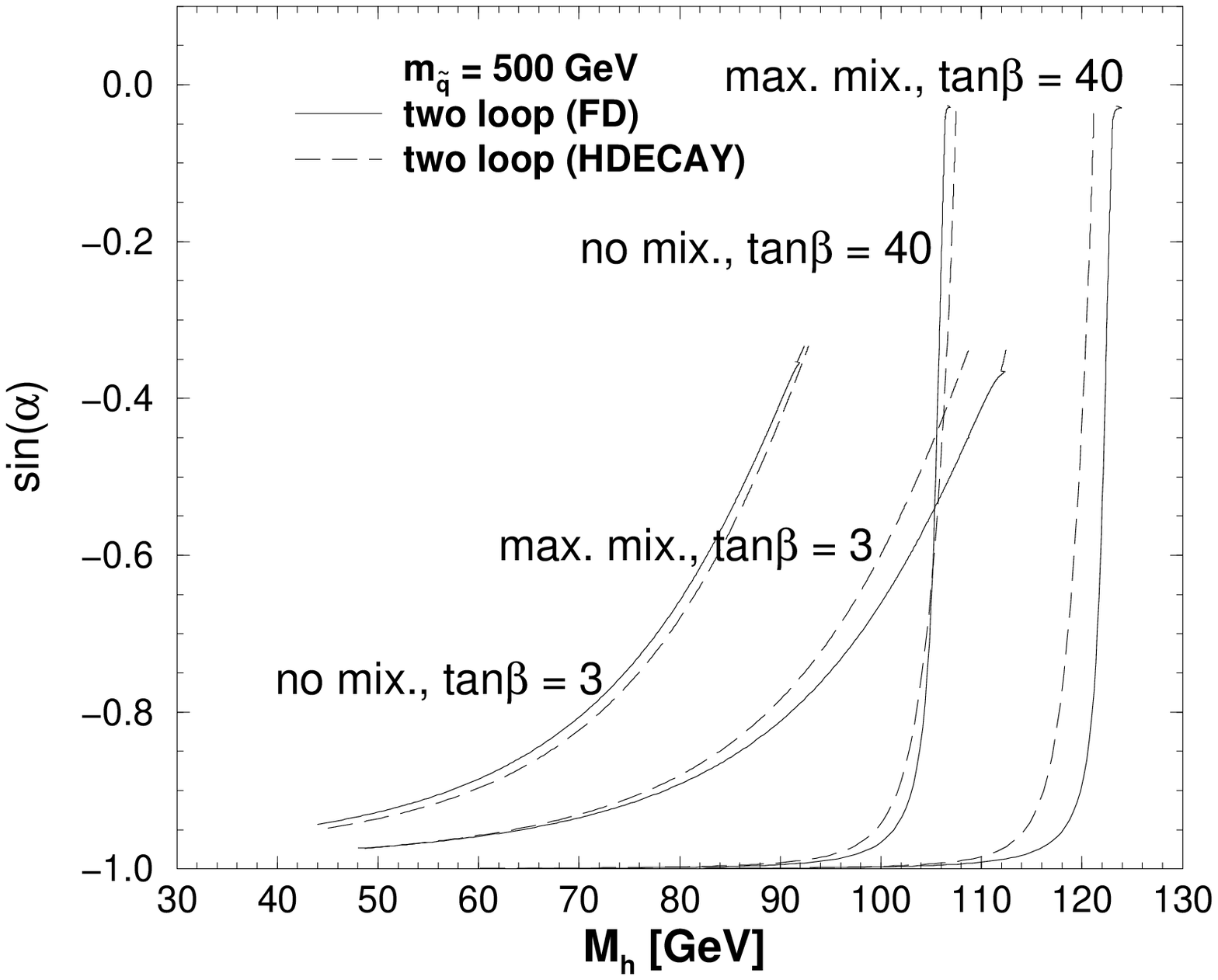,width=7cm,height=5.4cm}}
\hspace{1.5em}
\mbox{
\psfig{figure=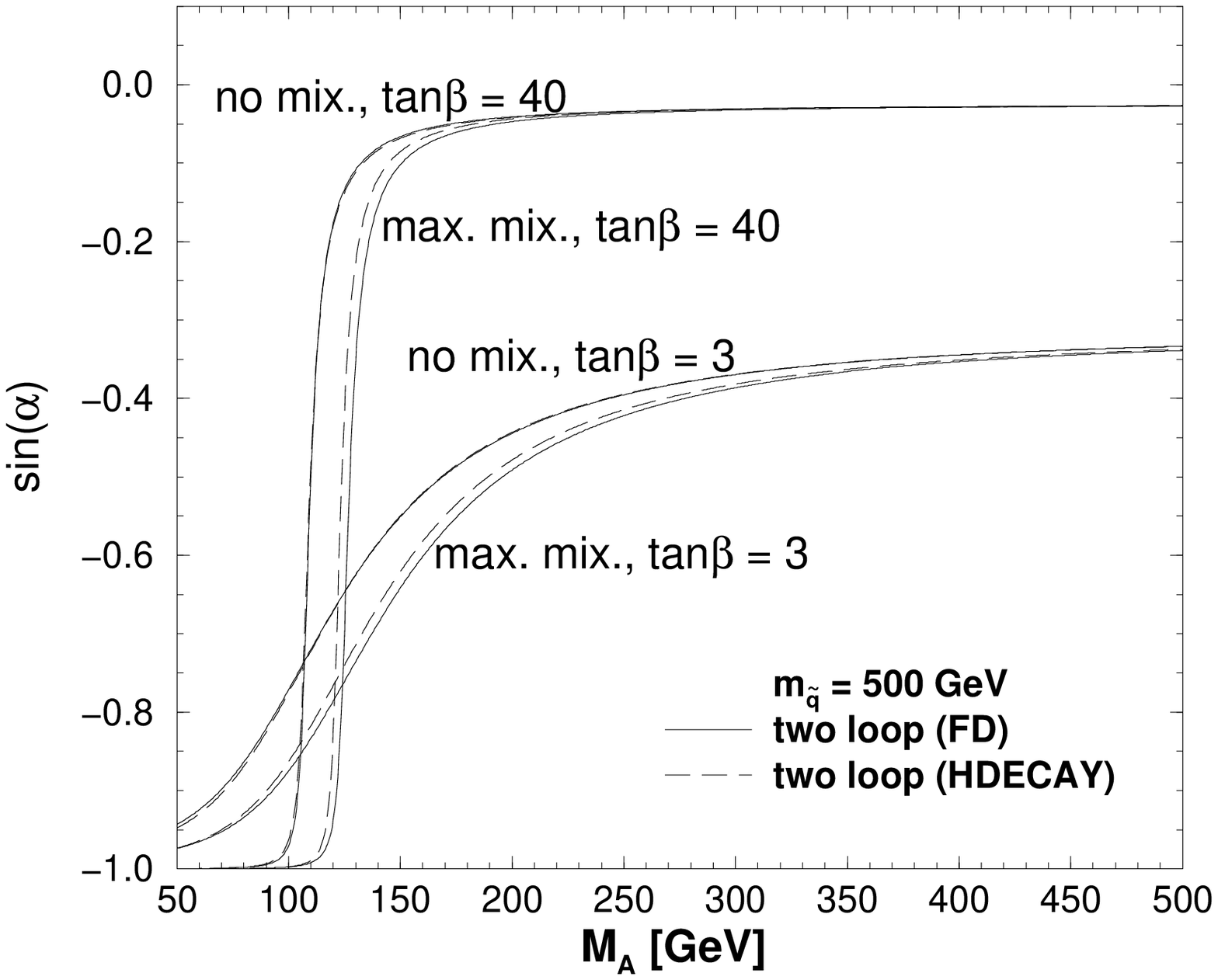,width=7cm,height=5.4cm}}
\end{center}
\caption[]{{\it
$\Ga(\hbb)$ is shown as a function of $\Mh$ in the left part and as a
function of $\MA$ in the right part. 
The results of the RG approach (without gluino contribution) are
compared with the Feynman diagrammatic results (without and with gluino
contribution.) The other parameters are
$\mu = -100 \gev$, $M_2 = \msq$, $\mgl = 500 \gev$, $\Ab = \At$.
In each plot the no-mixing and the
maximal-mixing scenarios are shown.
In the lower part the sine of the effective mixing angle is shown.
}}
\label{fig:Ghbbrgv}
\end{figure}

In order to compare our results with those obtained within the 
renormalization-group-improved effective field theory 
approach~\cite{mhiggsRG1a,mhiggsRG1b} (in the
following, for brevity reasons, denoted as RG 
approach), we made use of the program
HDECAY~\cite{hdecay}. In \reffi{fig:Ghbbrgv} we show the decay rate
$\Ga(\hbb)$ as a function of $\Mh$ (left part) and as a function of
$\MA$ (right part). 
We compare the RG result of HDECAY, where the
gluino-exchange contribution is not yet implemented, with the
FD result without and with gluino
correction.
In order to trace back the source of deviations we also show in the
lower part of \reffi{fig:Ghbbrgv} the sine of the effective mixing
angle $\aeff$ which enters the decay rate $\Ga(\hbb)$ quadratically
(see \refeqs{ampeffhbb} and (\ref{gameff})). 
In order to evaluate $\aeff$ we have neglected the external momentum
(see \refeq{hseq2zero}).
As a typical soft SUSY-breaking scale we have
chosen $\msq = 500 \gev$ in \reffi{fig:Ghbbrgv}. Since the FD and the
RG result have been 
obtained in different renormalization schemes, the entries of the
$\Stop$-mass matrix have a different meaning starting from \twol\
order~\cite{mhiggslle,FDRG1,bse}. 
For the no-mixing case we have set
$\Xt = 0$ in both approaches, whereas the maximal-mixing case is defined via 
$\Xt = 2\,\msq$ for the FD approach and $\Xt = \sqrt{6}\,\msq$ in
the RG approach~\cite{mhiggslle}.

\begin{figure}[ht!]
\vspace{2em}
\begin{center}
\mbox{
\psfig{figure=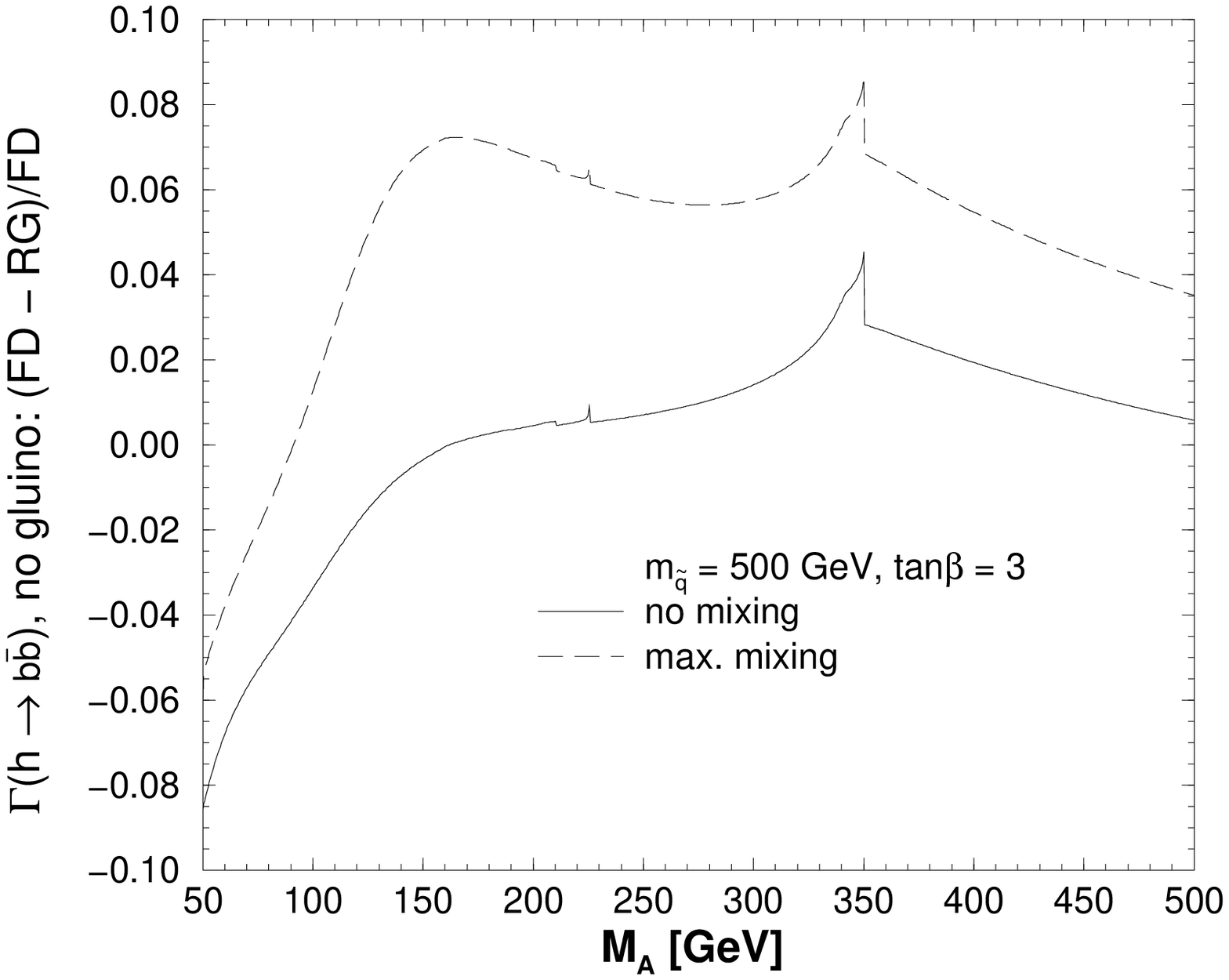,width=7cm,height=7cm}}
\hspace{1.5em}
\mbox{
\psfig{figure=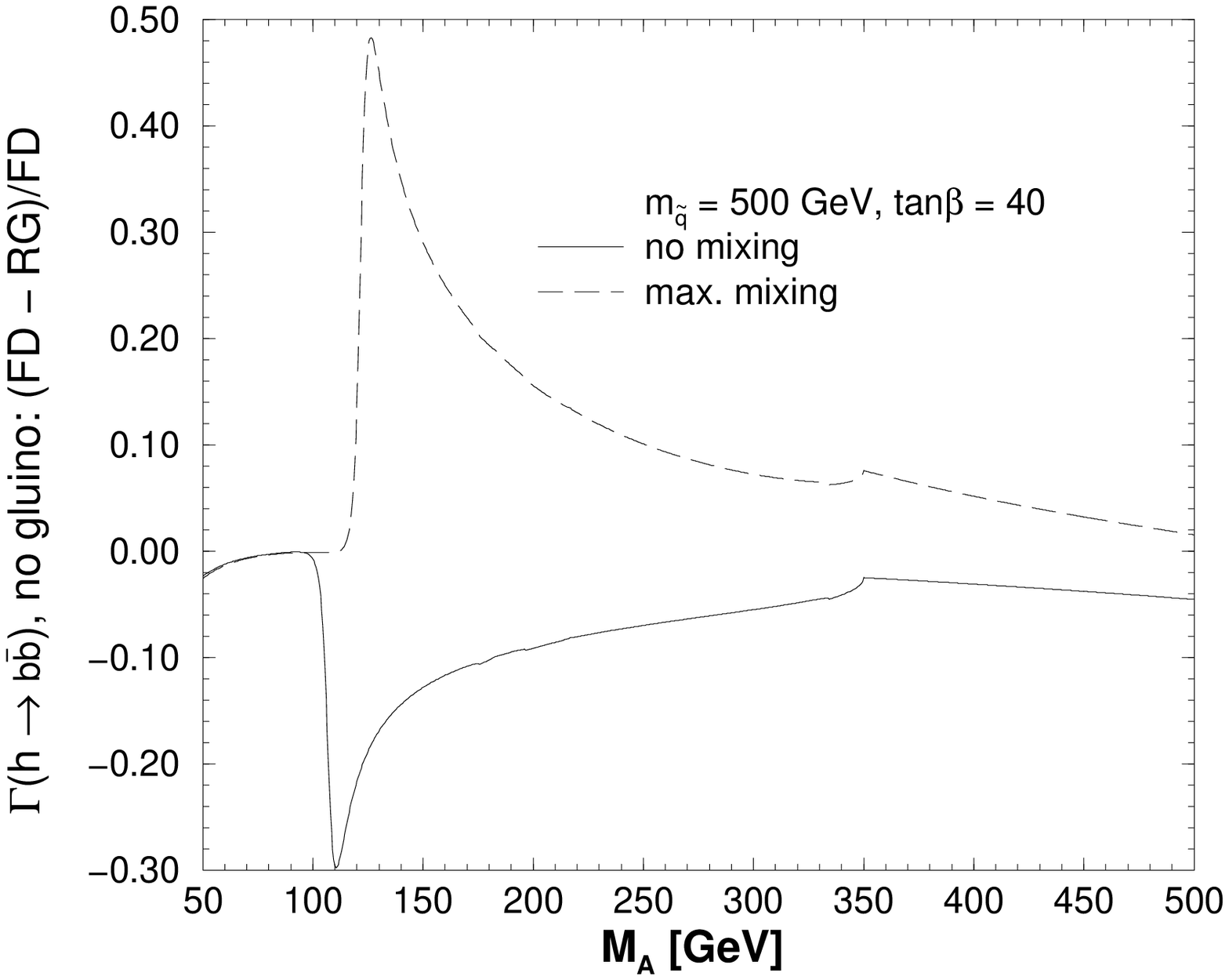,width=7cm,height=7cm}}
\end{center}
\caption[]{{\it
$\De\Ga(\hbb) = 
(\Ga^{\rm FD}(\hbb) - \Ga^{\rm RG}(\hbb))/\Ga^{\rm FD}(\hbb)$ 
is shown as a function of $\MA$ for no mixing and maximal mixing.
The gluino-contributions are neglected here. 
The other parameters are 
$\mu = -100 \gev$, $M_2 = \msq$, $\mgl = 500 \gev$, $\Ab = \At$.
The results are given in the low $\tb$ and in the high $\tb$ scenario.
}}
\label{fig:RCGhbbrgv}
\end{figure}
%
\begin{figure}[hb!]
\begin{center}
\mbox{
\psfig{figure=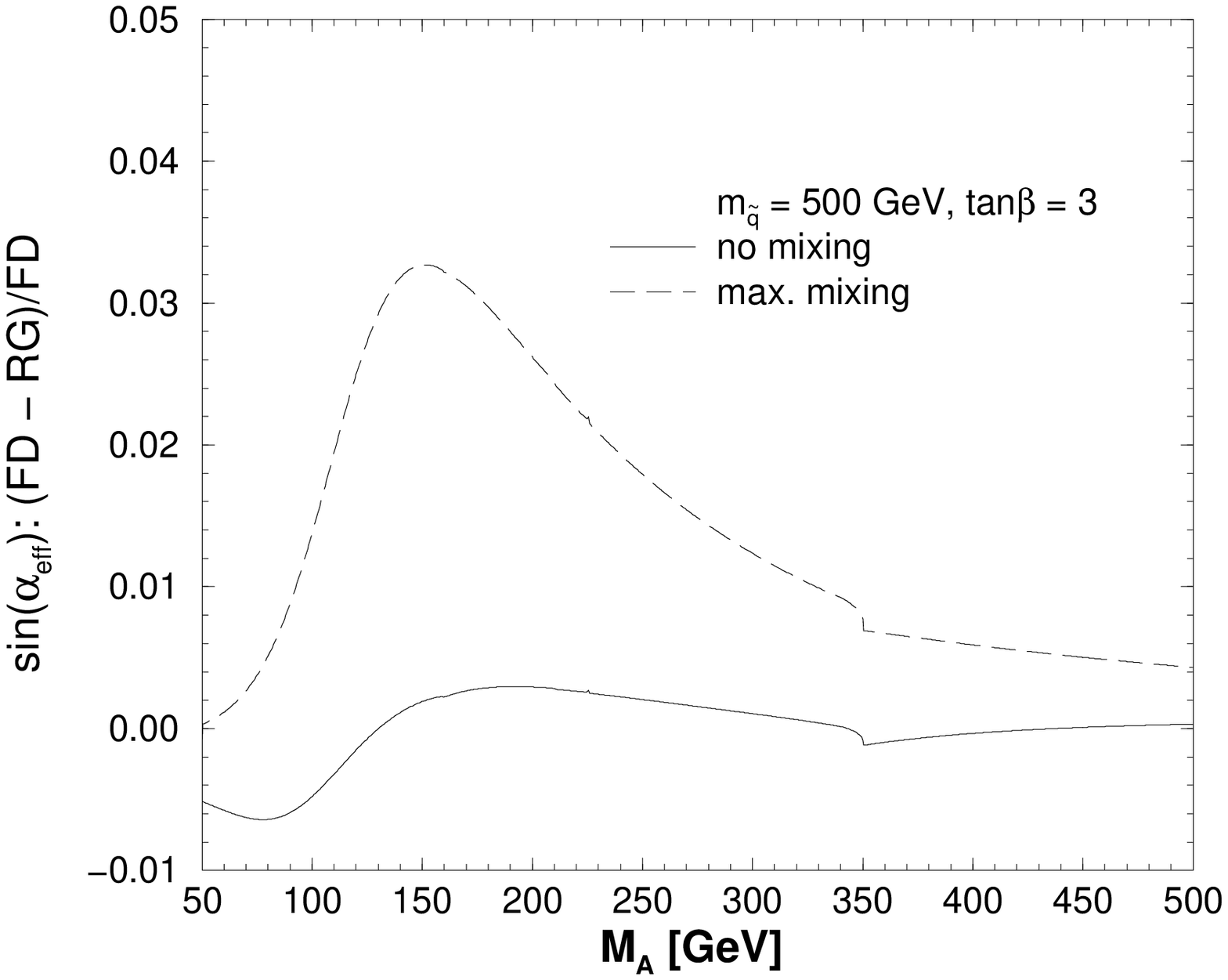,width=7cm,height=7cm}}
\hspace{1.5em}
\mbox{
\psfig{figure=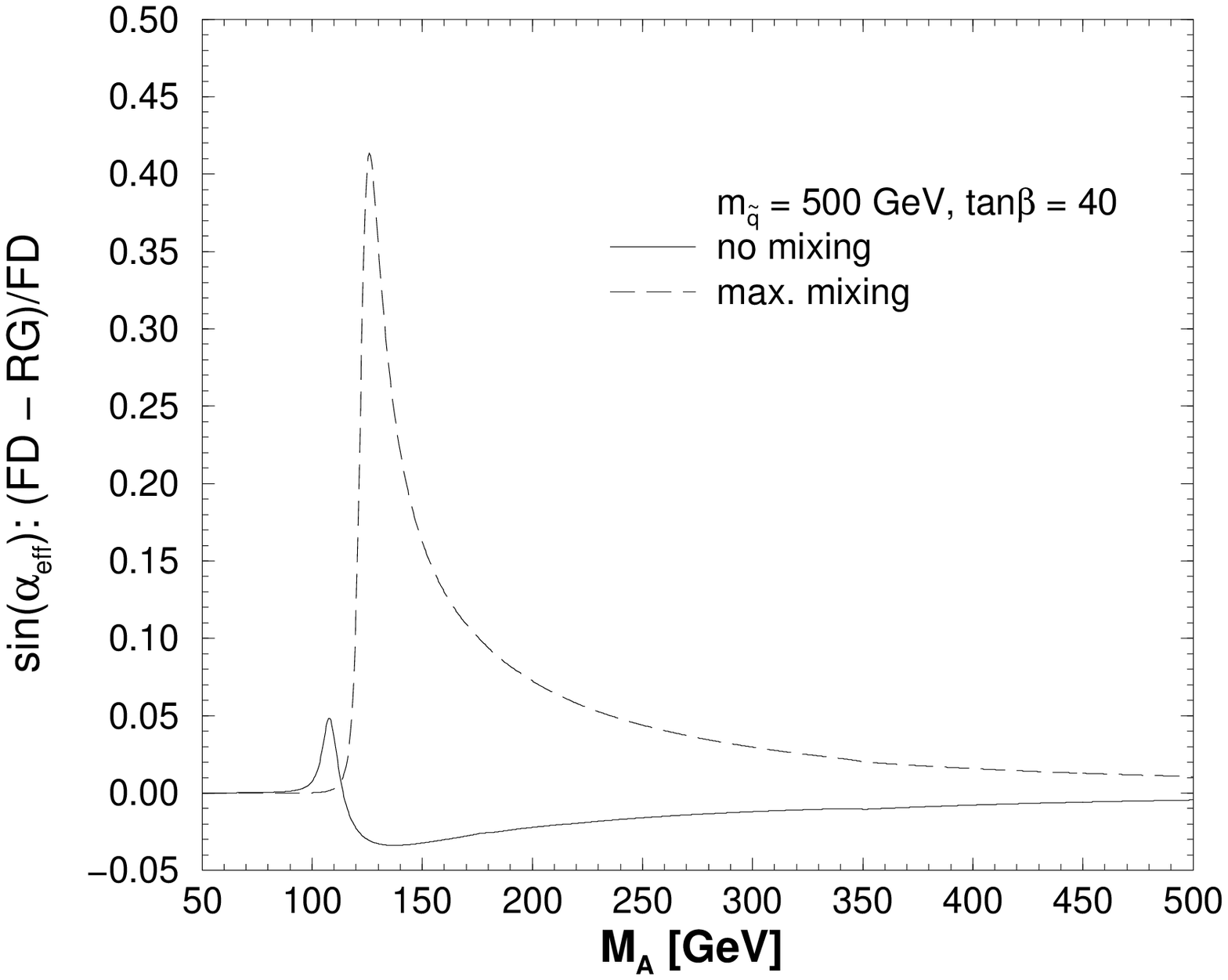,width=7cm,height=7cm}}
\end{center}
\caption[]{{\it
The relative difference 
$(\sin\aeff^{\rm FD} - \sin\aeff^{\rm RG})/\sin\aeff^{\rm FD}$  
is shown as a function
of $\MA$ for two values of $\tb$ in the no mixing and the maximal
mixing scenario. The other parameters are chosen as in
\reffi{fig:RCGhbbrgv}. 
}}
\label{fig:RCsaeffrgv}
\end{figure}

In the comparison of the decay rates (and restricting ourselves
to the results with neglected gluino exchange), for a given $\Mh$ we
find deviations for low $\tb$ up to ${\cal O}(10\%)$ and 
agreement better than 4\% in the large $\tb$ scenario.
The main part of the
deviations can be attributed to the deviations in $\aeff$
which modify $\Ga(\hbb)$ (see the lower part of \reffi{fig:Ghbbrgv}).
As a general feature, larger  
deviations arise at the endpoints of the $\Mh$ spectrum due
to the fact that for the same value of $\MA$ 
different values for $\Mh$ are obtained in the FD and the RG
approach, as shown in the left part
of \reffi{fig:Ghbbrgv}.
The plots for the $\Tb = 40$ scenario show again a sizable effect of
the gluino correction.

In \reffis{fig:RCGhbbrgv} and \reffi{fig:RCsaeffrgv} the relative
differences $\De\Ga(\hbb) = 
(\Ga^{\rm FD}(\hbb) - \Ga^{\rm RG}(\hbb))/\Ga^{\rm FD}(\hbb)$ 
and $(\sin\aeff^{\rm FD} - \sin\aeff^{\rm RG})/\sin\aeff^{\rm FD}$  
are shown as a function of $\MA$.
Comparing $\aeff$ as a function of $\MA$ the agreement is relatively
good; sizable deviations larger than 10\% are found only in the
regions where the curves in the right part of \reffi{fig:Ghbbrgv}
have a steep slope. This can give rise to large deviations up to 50\%
in $\Ga(\hbb)$ in terms of $\MA$ (for $100 \gev \lsim \MA \lsim 150 \gev$), 
as displayed in the right parts of \reffis{fig:RCGhbbrgv} and
\ref{fig:RCsaeffrgv} (large $\tb$).



\section{Conclusions}

Using the Feynman-diagrammatic approach for the Higgs-boson
propagator corrections, including besides the full \onel\ result also
the dominant \twol\ corrections, we have calculated the decay
rates and branching ratios for the decays $\hbb$, $\hcc$ and
$\htautau$. We have included the \onel\ QED and QCD 
corrections, where the latter are due to 
gluon and gluino-exchange contributions.
The gluino-exchange corrections have been neglected in
most of the previous phenomenological analyses. 
In the present analysis, only the purely weak $\oa$ (process specific)
vertex corrections, shown to contribute less 
than 1\% for most parts of the MSSM parameter space, have been neglected.

We have shown analytically that the full set of Higgs-boson propagator
corrections for vanishing external momentum can be absorbed into the
effective mixing angle, $\aeff$, in the neutral $\cp$-even Higgs
sector, appearing in the Higgs-boson fermion couplings.

Numerically we have shown that, compared in terms of $\Mh$, the \twol\
contributions to the 
Higgs-boson propagator corrections lead to a sizable decrease for
$\Ga(\hbb)$ and $\Ga(\htautau)$, whereas $\Ga(\hcc)$ can be increased,
although it stays relatively small for all sets of parameters we have
investigated. 
A sizable difference in all analyses from the one- to the \twol\
calculation arises from the kinematical effect that at the \twol\
level lower values for $\Mh$ are obtained compared to the \onel\ case,
thus leading to deviations 
at the endpoints of the shown $\Mh$ spectra.
For most parts of the MSSM parameter space the $\aeff$-approximation
reproduces the full result better than 3\%.
The gluino-exchange contribution to $\Ga(\hbb)$ has been
shown to be sizable for large $\tb$. This correction
increases with rising $\mgl$ and decreases with rising $\msq$ and
$\MA$. It is 
positive (negative) for negative (positive) $\mu$. In the $\tb = 40$
scenario for small $\msq$ it nearly compensate the gluon-exchange
correction. 

The effect of the Higgs-propagator corrections on the branching ratios
$BR(\hff)$ is relatively small for small values of $\Mh$, while
sizable effects are possible in the experimentally favored region of
$\Mh$. The branching ratio $BR(\htautau)$ can receive large
corrections up to 50\% due to SUSY-QCD corrections to $\Ga(\hbb)$.
In some parameter regions the effective $hb\bar{b}$ coupling can
become very small, and $BR(\hbb)$ can approach zero. In these parameter
regions the \twol\ contributions as well as the effect of the momentum
dependence of the \onel\ contributions are particularly important.
It should be noted that in order to determine $BR(\hbb)$ very precisely
in these paremeter regions,
also a resummation of the leading terms as well as the inclusion of the
complete $\oa$ vertex corrections will be necessary.

We have compared our diagrammatic result for $\Ga(\hbb)$ with the
result obtained within the renormalization-group-improved effective
field theory approach.
For the low $\tb$ scenario, compared in terms of $\Mh$,  we find
deviations up to ${\cal O}(10\%)$. 
In the large $\tb$ scenario the agreement is better than 4\%.
Compared in terms of $\MA$, however, differences of up to 50\% are
possible in the region $100 \gev \lsim \MA \lsim 150 \gev$.
The main part of the deviations can be
attributed to the differences in the effective mixing angle $\aeff$.


\bigskip
\subsection*{Acknowledgments}

We thank M.~Spira for valuable discussions and communication about the
numerical comparison of our results. We are also grateful to M.~Carena,
H.~Eberl and C.~Wagner for interesting and helpful discussions.
Parts of the calculations have been performed on the QCM cluster at the
University of Karlsruhe. 





\end{document}

%% file: draft.tex
\def\draftdate{\relax}
\def\mda{\relax}
\def\mua{\relax}
\def\mla{\relax}
\def\draft{
\def\thtystars{******************************}
\def\sixtystars{\thtystars\thtystars}
\typeout{}
\typeout{\sixtystars**}
\typeout{* Draft mode!
         For final version remove \protect\draft\space in source file *}
\typeout{\sixtystars**}
\typeout{}
\def\draftdate{\today}
\def\mua{\marginpar[\boldmath\hfil$\uparrow$]%
                   {\boldmath$\uparrow$\hfil}%
                    \typeout{marginpar: $\uparrow$}\ignorespaces}
\def\mda{\marginpar[\boldmath\hfil$\downarrow$]%
                   {\boldmath$\downarrow$\hfil}%
                    \typeout{marginpar: $\downarrow$}\ignorespaces}
\def\mla{\marginpar[\boldmath\hfil$\rightarrow$]%
                   {\boldmath$\leftarrow $\hfil}%
                    \typeout{marginpar: $\leftrightarrow$}\ignorespaces}
\def\Mua{\marginpar[\boldmath\hfil$\Uparrow$]%
                   {\boldmath$\Uparrow$\hfil}%
                    \typeout{marginpar: $\Uparrow$}\ignorespaces}
\def\Mda{\marginpar[\boldmath\hfil$\Downarrow$]%
                   {\boldmath$\Downarrow$\hfil}%
                    \typeout{marginpar: $\Downarrow$}\ignorespaces}
\def\Mla{\marginpar[\boldmath\hfil$\Rightarrow$]%
                   {\boldmath$\Leftarrow $\hfil}%
                    \typeout{marginpar: $\Leftrightarrow$}\ignorespaces}
\overfullrule 5pt
\oddsidemargin -15mm
\evensidemargin -15mm
\marginparwidth 29mm
}